\documentclass[twocolumn, floatfix, tighten, twocolappendix]{aastex63}

\usepackage[T1]{fontenc}
\usepackage{xcolor}
\usepackage{ulem}
 \usepackage{mathrsfs}

\makeatletter
\DeclareRobustCommand{\HI}{%
  \mbox{H\check@mathfonts\fontsize\sf@size\z@\selectfont I}%
}
\makeatother
 
\newcommand{\lya}{Ly$\alpha$}
\newcommand{\lyb}{Ly${\beta}$}
\newcommand{\lyg}{Ly$\gamma$}

\newcommand{\rom}[1]{\uppercase\expandafter{\romannumeral #1\relax}}

\def\cmpch{{h^{-1}{\rm Mpc}}} %
\def\teff{\tau_{\rm eff} }
\def\zem{z_{\rm em}}

\def\xhi{\langle x_{\rm HI}\rangle}

\def\cut#1{{\color{black}{}\color{black}}}

\def\add#1{{\color{black}{#1}\color{black}}}

\defcitealias{zhu_chasing_2021}{Paper I}

\shorttitle{Dark Gaps in the Ly$\beta$ Forest}
\shortauthors{Zhu et al.}

\begin{document}

\title{Long Dark Gaps in the Ly$\beta$ Forest at $z<6$: Evidence of Ultra Late Reionization from XQR-30 Spectra
}

\author[0000-0003-3307-7525]{Yongda Zhu}
\affiliation{Department of Physics \& Astronomy,
    University of California, Riverside, CA 92521, USA;
    \href{mailto:yzhu144@ucr.edu}{\emph{yzhu144@ucr.edu}}
    }

\author[0000-0003-2344-263X]{George D. Becker}
\affiliation{Department of Physics \& Astronomy,
    University of California, Riverside, CA 92521, USA;
    \href{mailto:yzhu144@ucr.edu}{\emph{yzhu144@ucr.edu}}
    }

\author[0000-0001-8582-7012]{Sarah E. I. Bosman}
\affiliation{Max-Planck-Institut f\"{u}r Astronomie, K\"{o}nigstuhl 17, D-69117 Heidelberg, Germany}

\author[0000-0001-5211-1958]{Laura C. Keating}
\affiliation{Leibniz-Institut f\"ur Astrophysik Potsdam (AIP), An der Sternwarte 16, D-14482 Potsdam, Germany}

\author[0000-0003-3693-3091]{Valentina D'Odorico}
\affiliation{INAF-Osservatorio Astronomico di Trieste, Via Tiepolo 11, I-34143 Trieste, Italy}
\affiliation{Scuola Normale Superiore, Piazza dei Cavalieri 7, I-56126 Pisa, Italy}
\affiliation{IFPU-Institute for Fundamental Physics of the Universe, via Beirut 2, I-34151 Trieste, Italy}

\author[0000-0002-3324-4824]{Rebecca L. Davies}
\affiliation{Centre for Astrophysics and Supercomputing, Swinburne University of Technology, Hawthorn, Victoria 3122, Australia}
\affiliation{ARC Centre of Excellence for All Sky Astrophysics in 3 Dimensions (ASTRO 3D), Australia}

\author[0000-0002-0421-065X]{Holly M. Christenson}
\affiliation{Department of Physics \& Astronomy,
    University of California, Riverside, CA 92521, USA;
    \href{mailto:yzhu144@ucr.edu}{\emph{yzhu144@ucr.edu}}
    }
\author[0000-0002-2931-7824]{Eduardo Ba\~nados}
\affiliation{Max-Planck-Institut f\"{u}r Astronomie, K\"{o}nigstuhl 17, D-69117 Heidelberg, Germany}

\author[0000-0002-1620-0897]{Fuyan Bian}
\affiliation{European Southern Observatory, Alonso de Córdova 3107, Casilla 19001, Vitacura, Santiago 19, Chile}

\author[0000-0002-4314-021X]{Manuela Bischetti}
\affiliation{INAF-Osservatorio Astronomico di Trieste, Via Tiepolo 11, I-34143 Trieste, Italy}

\author[0000-0002-3211-9642]{Huanqing Chen}
\affiliation{Department of Astronomy \& Astrophysics; 
The University of Chicago; 
Chicago, IL 60637, USA}

\author[0000-0003-0821-3644]{Frederick B.~Davies}
\affiliation{Max-Planck-Institut f\"{u}r Astronomie, K\"{o}nigstuhl 17, D-69117 Heidelberg, Germany}

\author[0000-0003-2895-6218]{Anna-Christina Eilers}\thanks{NASA Hubble Fellow}
\affiliation{MIT Kavli Institute for Astrophysics and Space Research, 77 Massachusetts Avenue, Cambridge, MA 02139, USA}

\author[0000-0003-3310-0131]{Xiaohui Fan}
\affiliation{Steward Observatory, University of Arizona, 933 North Cherry Avenue, Tucson, AZ 85721, USA}

\author[0000-0002-2423-7905]{Prakash Gaikwad}
\affiliation{Max-Planck-Institut f\"{u}r Astronomie, K\"{o}nigstuhl 17, D-69117 Heidelberg, Germany}

\author[0000-0002-4085-2094]{Bradley Greig}
\affiliation{School of Physics, University of Melbourne, Parkville, VIC 3010, Australia}
\affiliation{ARC Centre of Excellence for All Sky Astrophysics in 3 Dimensions (ASTRO 3D), Australia}

\author[0000-0001-8443-2393]{Martin G. Haehnelt}
\affiliation{Kavli Institute for Cosmology and Institute of Astronomy, Madingley Road, Cambridge, CB3 0HA, UK}

\author[0000-0001-5829-4716]{Girish Kulkarni}
\affiliation{Tata Institute of Fundamental Research, Homi Bhabha Road, Mumbai 400005, India}

\author[0000-0001-9372-4611]{Samuel Lai}
\affiliation{Research School of Astronomy and Astrophysics, Australian National University, Canberra, ACT 2611, Australia}

\author[0000-0002-7129-5761]{Andrea Pallottini}
\affiliation{Scuola Normale Superiore, Piazza dei Cavalieri 7, I-56126 Pisa, Italy}

\author[0000-0002-4314-1810]{Yuxiang Qin}
\affiliation{School of Physics, University of Melbourne, Parkville, VIC 3010, Australia}
\affiliation{ARC Centre of Excellence for All Sky Astrophysics in 3 Dimensions (ASTRO 3D), Australia}

\author[0000-0002-5360-8103]{Emma V. Ryan-Weber}
\affiliation{Centre for Astrophysics and Supercomputing, Swinburne University of Technology, Hawthorn, Victoria 3122, Australia}
\affiliation{ARC Centre of Excellence for All Sky Astrophysics in 3 Dimensions (ASTRO 3D), Australia}

\author[0000-0003-4793-7880]{Fabian Walter}
\affiliation{Max-Planck-Institut f\"{u}r Astronomie, K\"{o}nigstuhl 17, D-69117 Heidelberg, Germany}

\author[0000-0002-7633-431X]{Feige Wang}\thanks{NASA Hubble Fellow}
\affiliation{Steward Observatory, University of Arizona, 933 North Cherry Avenue, Tucson, AZ 85721, USA}

\author[0000-0001-5287-4242]{Jinyi Yang}\thanks{Strittmatter Fellow}
\affiliation{Steward Observatory, University of Arizona, 933 North Cherry Avenue, Tucson, AZ 85721, USA}

\begin{abstract}
    We present a new investigation of the intergalactic medium (IGM) near reionization
    using dark gaps in the Lyman-$\beta$ (\lyb) forest.  With its lower optical depth, \lyb\ offers a potentially more sensitive probe to any remaining neutral gas compared to commonly used \lya\ line. We identify dark gaps in the \lyb\ forest using spectra of 42 QSOs at $z_{\rm em}>5.5$, including new data from the XQR-30 VLT Large Programme. Approximately 40\% of these QSO spectra exhibit dark gaps longer than $10\cmpch$ at $z\simeq5.8$. By comparing the results to predictions from simulations, we find that the data are broadly consistent both with models where fluctuations in the \lya\ forest are caused solely by ionizing ultraviolet background (UVB) fluctuations and with models that include large neutral hydrogen patches at $z<6$ due to a late end to reionization. 
    Of particular interest is a very long ($L=28\cmpch$) and dark ($\teff \gtrsim 6$) gap persisting down to $z\simeq 5.5$ in the \lyb\ forest of the $z_{\rm}=5.85$ QSO PSO J025$-$11. This gap may support late reionization models with a volume-weighted average neutral hydrogen fraction of
    $ \xhi \gtrsim 5\%$ by $z=5.6$.  Finally, we infer constraints on $\xhi$ over $5.5 \lesssim z \lesssim 6.0$ based on the observed \lyb\ dark gap length distribution and a conservative relationship between gap length  and neutral fraction derived from simulations. We find $\xhi \leq 0.05$, 0.17, and 0.29 at $z\simeq 5.55$, 5.75, and 5.95, respectively.  These constraints are consistent with models where reionization ends significantly later than $z = 6$.

\end{abstract}

\keywords{\href{http://astrothesaurus.org/uat/1383}{Reionization (1383)}, 
\href{http://astrothesaurus.org/uat/813}{Intergalactic medium (813)}, 
\href{http://astrothesaurus.org/uat/1317}{Quasar absorption line spectroscopy (1317)}, \href{http://astrothesaurus.org/uat/734}{High-redshift galaxies (734)}
}

\section{Introduction \label{sec:introduction}}

Determining when and how reionization occurred is essential for understanding the IGM physics and galaxy formation in the early Universe \citep[e.g.,][]{munoz_impact_2022}. Recent observations have made significant progress on establishing the timing of reionization and largely point to a midpoint of $z\sim 7-8$. These observations include the electron optical depth to the cosmic microwave background (CMB) photons \citep[][see also \citealp{de_belsunce_inference_2021}]{planck_collaboration_planck_2020}, the Lyman-$\alpha$ (\lya) damping wing in $z\gtrsim7$ QSO spectra \citep[e.g.,][]{banados_800-million-solar-mass_2018,davies_quantitative_2018,wang_significantly_2020-1,yang_poniuaena_2020,greig_igm_2021-1}, the decline in observed \lya\ emission from $z>6$ galaxies \citep[e.g.,][and references therein, but see \citealp{jung_texas_2020-1,wold_lager_2021-1}]{mason_universe_2018,mason_inferences_2019,hoag_constraining_2019,hu_ly_2019}, and the IGM thermal state measurements at $z>5$ \citep[e.g.,][]{boera_revealing_2019,gaikwad_consistent_2021}. 

The observations noted above are generally consistent with reionization ending by $z\sim6$, a scenario supported by existing measurement of the fraction of dark pixels in the Lyman series forests \citep[e.g.,][]{mcgreer_first_2011,mcgreer_model-independent_2015}.  Other observations, however, 
suggest a significantly later end of reionization. The large-scale fluctuations in the measured \lya\ effective optical depth, $\teff=-\ln{\langle F \rangle}$, where $F$ is the continuum-normalized transmission flux \citep[e.g.,][]{fan_constraining_2006,becker_evidence_2015,eilers_opacity_2018,bosman_hydrogen_2021,yang_measurements_2020-1}, together with long troughs extending to or below $z\simeq5.5$ in the \lya\ forest \citep[e.g.,][]{becker_evidence_2015,zhu_chasing_2021} potentially indicate the existence of large neutral IGM islands \citep[e.g.,][]{kulkarni_large_2019,keating_long_2020,nasir_observing_2020}.  The underdensities around long dark gaps traced by \lya\ emitting galaxies (LAEs) are also consistent with a late reionization model wherein reionization ends at $z<6$ \citep[][]{becker_evidence_2018,kashino_evidence_2020, christenson_constraints_2021}. 

These \lya\ forest and LAE results are potentially consistent with an alternative scenario wherein the IGM is ionized by $z=6$ but retains large-scale fluctuations in the ionizing UV background down to lower redshifts \citep[][]{davies_determining_2018}. 
On the other hand, recent measurements of the mean free path of ionizing photons measured at $z=5.1$ and 6.0 \citep{becker_mean_2021} are difficult to reconcile with models wherein reionization ends at $z > 6$, and may instead prefer models wherein the IGM is still $20\%$ neutral or more at $z=6$ \citep[][]{becker_mean_2021,cain_short_2021,davies_predicament_2021}.
In addition, a reionization ending at $z < 6$ is consistent with models that reproduce a variety of observations \citep[e.g.,][]{weinberger_modelling_2019,choudhury_studying_2021, qin_reionization_2021}. %

One way of searching for signatures of late ($z_{\rm end} < 6$) reionization in the \lya\ forest is by investigating dark gaps, i.e., contiguous regions of strong absorption  \citep[e.g.,][]{songaila_approaching_2002,furlanetto_constraining_2004,paschos_statistical_2005,fan_constraining_2006,gallerani_glimpsing_2008,gnedin_cosmic_2017,nasir_observing_2020}. In \citet[][hereafter \citetalias{zhu_chasing_2021}]{zhu_chasing_2021}, we explored long dark gaps in the \lya\ forest and found that a fully ionized IGM with a homogeneous UVB is strongly ruled out down to $z\simeq5.3$. In contrast, late reionization models and a model wherein reionization ends by $z\sim6$ but retains large-scale UVB fluctuations are  consistent with the observations. Predictions for the \lya\ dark gap statistics are similar between the two types of models.  This is largely because realistic late reionization models also include UVB fluctuations, which are often associated with the neutral islands.  \lya\ also tends to saturate at a relatively low ($x_{\rm HI} \sim 10^{-3}$) neutral fraction, limiting its sensitivity to neutral gas. 

Given its lower optical depth \footnote{\lyb\ absorption has a shorter wavelength ($\lambda_{\rm Ly \beta} = 1025.72$\,\AA) and a lower oscillator strength ($f_{\rm Ly \beta} = 0.0791$) compared to those of \lya\ absorption ($\lambda_{\rm Ly \alpha} = 1215.67$\,\AA, $f_{\rm Ly \alpha} = 0.4164$). The ratio of optical depth is given by $\tau_{\rm Ly\beta}/\tau_{\rm Ly\alpha}=(f_{\rm Ly \beta} \lambda_{\rm Ly\beta}) / (f_{\rm Ly\alpha} \lambda_{\rm Ly\alpha})\simeq 0.16$. }, \lyb\ should be a more sensitive probe of neutral gas in the $z \lesssim 6$ IGM. As a result, ultra-late reionization models wherein neutral islands persist down to $z<5.5$ may produce more long \lyb\ dark gaps than can be produced by UVB fluctuations alone. Based on this feature, we can potentially use dark gaps in the \lyb\ forest to place stronger constraints on the timing of reionization and distinguish the late reionization models from the pure fluctuating UVB models. As presented in \citet{nasir_observing_2020}, distributions of dark gaps in the \lyb\ forest differ between these models most strongly on scales of $L\gtrsim10\cmpch$. We are therefore particularly interested in these long dark gaps. 

In this work, we use 42 high-quality QSO spectra that allow us to search for dark gaps in the \lyb\ forest over the redshift range $5.5 < z < 6.0$.  The sample includes high-quality X-Shooter spectra from the XQR-30 VLT large program \footnote{\href{https://xqr30.inaf.it}{https://xqr30.inaf.it}} (D'Odorico et al., in prep.). In addition to comparing the results to model predictions, we also constrain $ \langle x_{\rm HI} \rangle $ based on a conservative relationship between dark gap length and neutral fraction derived from simulations.  

This paper is organized as follows. In {Section~\ref{sec:data}} we describe the data and results from the observations. \mbox{Section~\ref{sec:models}} compares our results to model predictions, discusses the implications for reionization, and infers constraints on $x_{\rm HI}$.  Finally, we conclude the findings in \mbox{Section~\ref{sec:conslusion}}. Throughout this paper, we assume a $\Lambda$CDM cosmology with $\Omega_{\rm m}=0.308$, $\Omega_{\Lambda}=0.692$, and $h=0.678$ \citep[][]{planck_collaboration_planck_2014}. Distances are quoted in comoving units unless otherwise noted.

\section{The Data and Results \label{sec:data}}

\subsection{QSO Spectra}

\begin{figure*}[!t]
    \centering \includegraphics[width=7in]{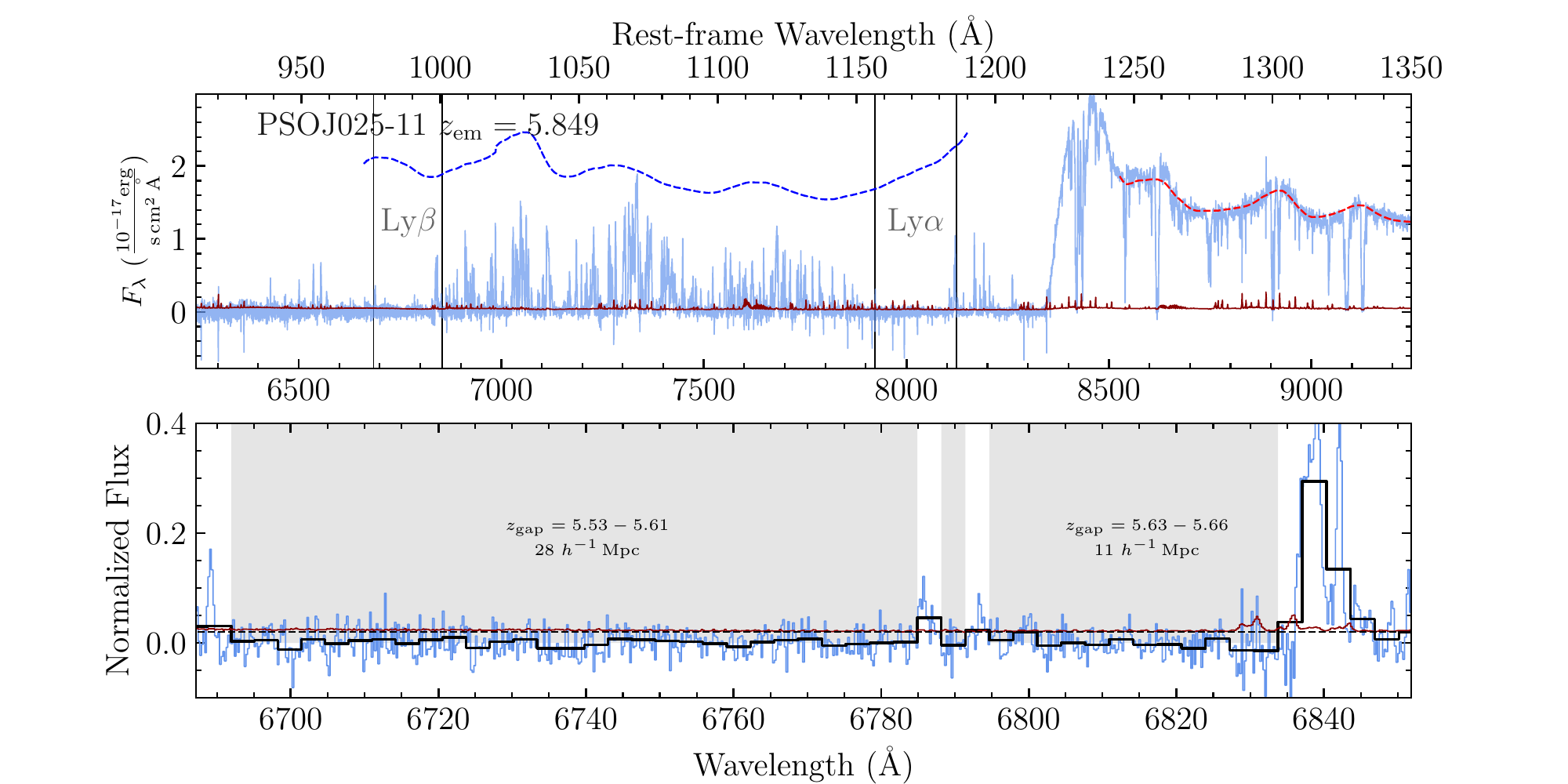}
    \caption{
        \textbf{{Top panel:}} 
        Spectrum and continuum fits for the $\zem = 5.849$ QSO PSO J025$-$11. 
        The light blue and dark red lines represents flux and flux error in the original binning. Dashed curves redward and blueward of the \lya\ peak show the best-fitting and predicted QSO continuum in absence of absorption, respectively. The fitting and prediction are based on Principal Component Analysis (PCA). The continuity between the \lya\ forest and the \lyb\ forest continuum is broken intentionally near 1020 \AA\ in the rest frame. We label the wavelength range over which we search for dark gaps in the \lyb\ forest and its corresponding \lya\ forest in redshift.
        \textbf{{Bottom panel:}} 
        \lyb\ forest and dark gaps detected. The dashed black line labels the flux threshold of 0.02. The thick black line displays the flux binned to $1\cmpch$. Light blue and dark red lines show the flux and flux error in the original binning. Dark gaps detected are shaded with gray. We also label the redshift range and length of each long ($L\geq10\cmpch$) dark gap. \\
        (The complete figure set [42 images] is available online at \url{https://ydzhuastro.github.io/lyb.html}.)
        }
        \label{fig:spectra}
\end{figure*}

Our sample includes the 42 out of 43 spectra of QSOs at $5.77 \lesssim z_{\rm em} \lesssim 6.31$ that were used for \citetalias{zhu_chasing_2021}.  The exception is PSO J004$+$17, whose spectrum has lower S/N that does not meet the requirement of our flux threshold for the \lyb\ forest (Section \ref{sec:detection-method}).  The spectra are taken with the Echellette Spectrograph and Imager (ESI) on Keck \citep{sheinis_esi_2002} and the X-Shooter spectrograph on the Very Large Telescope (VLT; \citealp{vernet_x-shooter_2011}). Among these, 19 X-Shooter spectra are from the XQR-30 VLT large program (D'Odorico et al., in prep). The details of the data reduction procedures are given in \citetalias{zhu_chasing_2021} and \citet{becker_evolution_2019}. We note that the targets are selected without foreknowledge of dark gaps in the \lyb\ forest. Figure Set~\ref{fig:spectra} displays the spectra, continuum fits, and dark gaps detected in the \lyb\ forest for each QSO (see details below).  An example is given in Figure~\ref{fig:spectra}.

\subsection{Continuum Fitting}
We use Principal Component Analysis (PCA) to predict the unabsorbed QSO continuum and normalize the transmission in the \lyb\ forest. We follow a similar method as described in \citetalias{zhu_chasing_2021} to fit and predict the continuum. Briefly, we implement and apply the log-PCA method of \cite{davies_predicting_2018} in the \lya\ and \lyb\ forest portion of the spectrum following \citet{bosman_hydrogen_2021}. The continuity between the \lya\ forest and the \lyb\ forest continuum is broken intentionally to correct for the effect of overlapping \lya\ absorption in the \lyb\ forest in the PCA training sample. We fit the red-side (rest-frame wavelength $\lambda_0>1230$ \AA) continuum up to 2000 \AA\ in the rest frame for X-Shooter spectra with NIR observations. The ESI spectra are fit using an optical-only PCA, which is presented in \citet{bosman_comparison_2021}. The \lyb\ dark gap detection is not very sensitive to the continuum, and we also test that using a power-law continuum does not significantly change the dark gap results in this work.

\subsection{Dark Gap Detection \label{sec:detection-method}}
Similar to the definition of a dark gap in the \lya\ forest in \citetalias{zhu_chasing_2021}, we define a dark gap in the \lyb\ forest to be a continuous spectral region in which all pixels binned to $1\cmpch$ have an observed normalized flux $F = F_{\rm obs}/F_{\rm c} < 0.02$, where $F_{\rm obs}$ is the observed flux and $F_{\rm c}$ is the continuum flux. The minimum length of a dark gap is $1\cmpch$ following \citetalias{zhu_chasing_2021}. 
\add{A bin size of $1\cmpch$ enables us to retain significant transmission profiles while reducing the influence of very small  peaks on dark gap detection.  The precise bin size should have relatively little impact on our results provided that the observations and mock spectra are treated consistently. We have verified that using bin sizes of $0.5\cmpch$ or $1.5\cmpch$ does not change our major conclusions, although the lengths of some dark gaps would change.} 
A flux threshold of 0.02 is used here instead of 0.05, which we used for the \lya\ gaps, because spectra in this sub-sample have higher signal-to-noise (S/N) levels. In addition, the \lyb\ forest at the redshifts that we are interested in is less contaminated by sky lines than the \lya\ forest. We have tested that using a threshold of 0.05 will not change our results fundamentally, although the difference between the models (Section~\ref{sec:models}) may become less significant. 
In order to reduce false detections caused by foreground \lya\ absorption, we further require that all \lyb\ dark gaps correspond to \lya\ dark gaps as defined in \citetalias{zhu_chasing_2021} over the same redshifts for both the observed and mock (Section \ref{sec:mock_spectra}) spectra. That is to say, each $1\cmpch$ bin in the \lyb\ dark gap also has a normalized flux less than 0.05 in the \lya\ forest at the same redshift.
\footnote{Based on our test, whether requiring gaps to be also dark in the \lya\ forest or not only affects a small fraction of gaps and does not change the results in this paper significantly. Although this requirement may not remove all false detection, it partially avoids contamination from random foreground density fluctuations.}
For reference, we present the relationship between \lyb\ dark gaps and \lya\ dark gaps in Appendix \ref{app:Lyb_Lya}.

For each QSO sightline, dark gaps are detected from 976 \AA\ in the rest frame to 11 proper-Mpc blueward of the QSO's redshift, which corresponds to approximately 1000 \AA\ in the rest frame. The lower wavelength limit ensures that the detection is not affected by the \lyg\ absorption. We use the higher limit to avoid the QSO proximity zone transmission, and the cut is comparable to the choice in, e.g., \citet{bosman_comparison_2021}. Following \citetalias{zhu_chasing_2021}, we also exclude from the statistical analysis an additional $10\cmpch$ buffer zone blueward of the proximity zone cut. This allows the pixel at the red end of each sightline to intersect a {\em possible} long ($L\ge10\cmpch$) dark gap, and hence helps to mitigate potential truncation issues. \footnote{Without this additional buffer zone, it is possible that the $F_{10}$ (Section \ref{sec:gap_statistics}) is underestimated near the red end of a sightline, since there can exist otherwise $\ge10\cmpch$ gaps that are truncated by the edge or peaks in the proximity zone.}

To avoid the contamination from sky line subtraction residuals, we mask out $\pm 75~{\rm km\,s^{-1}}$ intervals of the spectra centered at sharp peaks in the flux error array when searching for dark gaps. The exception is that we do not mask transmission with a $>3\sigma$ detection. In addition, we make no attempts to correct for the effects of damped \lya\ systems (DLAs) or metal-enriched absorbers, although known systems in a sub-sample of the spectra with a relatively complete identification of metal-enriched systems are discussed in   \mbox{Appendix~\ref{app:moreMetal}}. Figure~\ref{fig:all_los} displays an overview of dark gaps detected in the \lyb\ forest from our sample. 

\begin{figure*}
    \centering
    \includegraphics[width=6in]{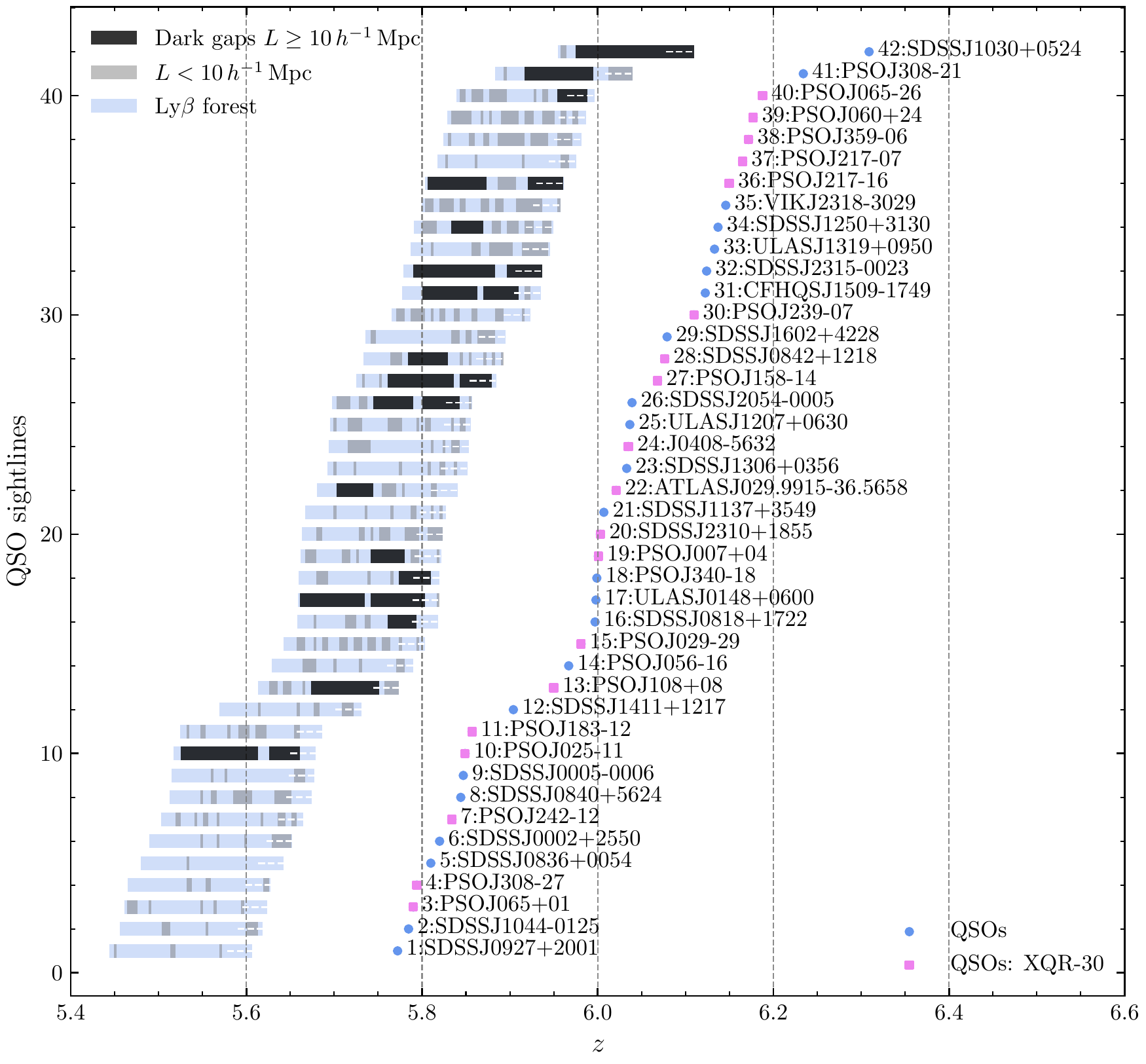}
    \caption{Overview of dark gaps identified in the \lyb\ forest from our sample of 42 QSO spectra. Black (gray) bars represent dark gaps longer (shorter) than $10\cmpch$. Pink squares label redshifts of XQR-30 targets, and blue dots mark the redshifts of the rest of QSOs.  Light blue shaded regions demonstrate the redshift coverage of the \lyb\ forest. We note that the \lyb\ forest is truncated at 11 pMpc from the QSO. The \lyb\ forest shown in this figure includes the $10\cmpch$ buffer zone labeled with dashed white line at the red end. 
    \label{fig:all_los}}
\end{figure*}

\subsubsection{Dark Gaps Statistics}\label{sec:gap_statistics}

\begin{figure*}
\centering
    \gridline{
    \hspace{-1cm}
        \fig{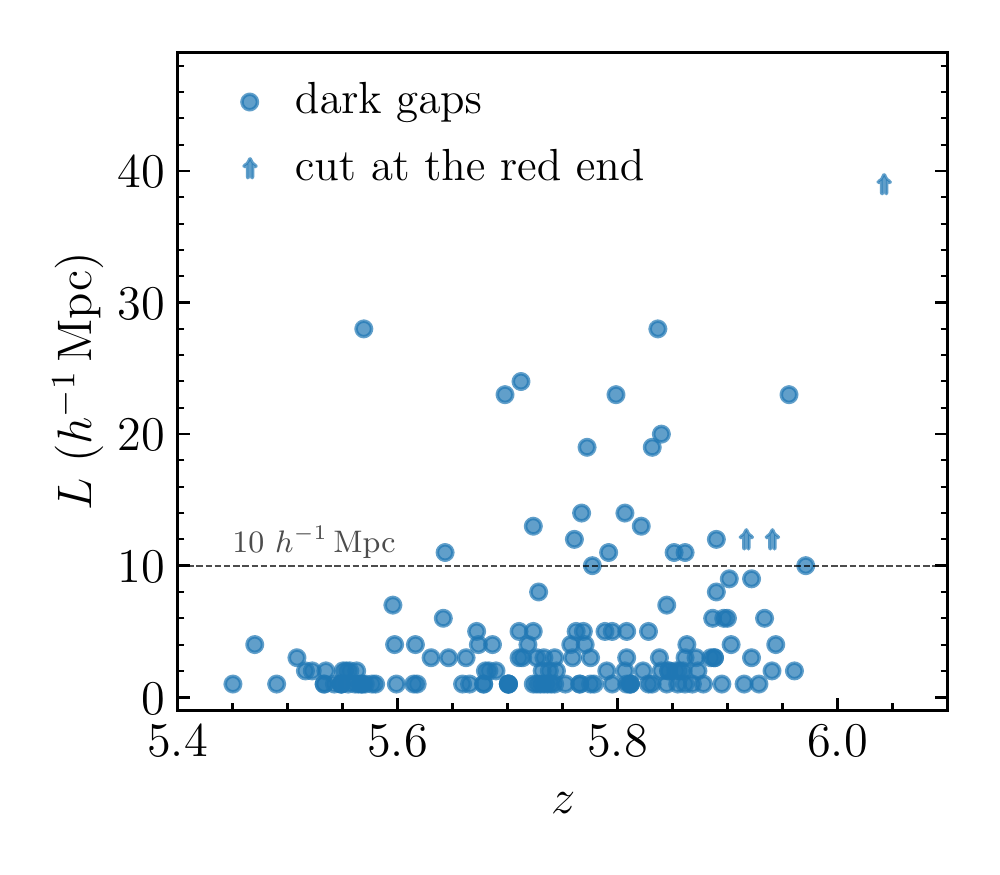}{2.6in}{(a)}
    \hspace{-0.3cm}
        \fig{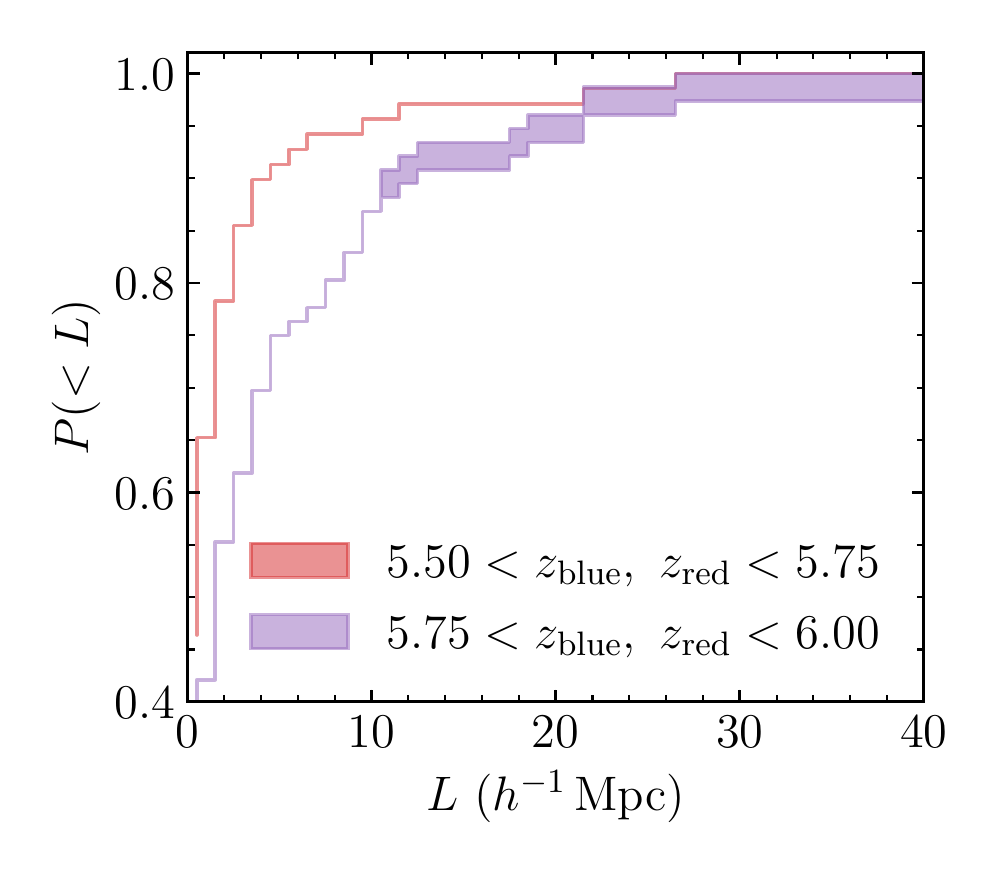}{2.6in}{(b)}
    \hspace{-0.4cm}
        \fig{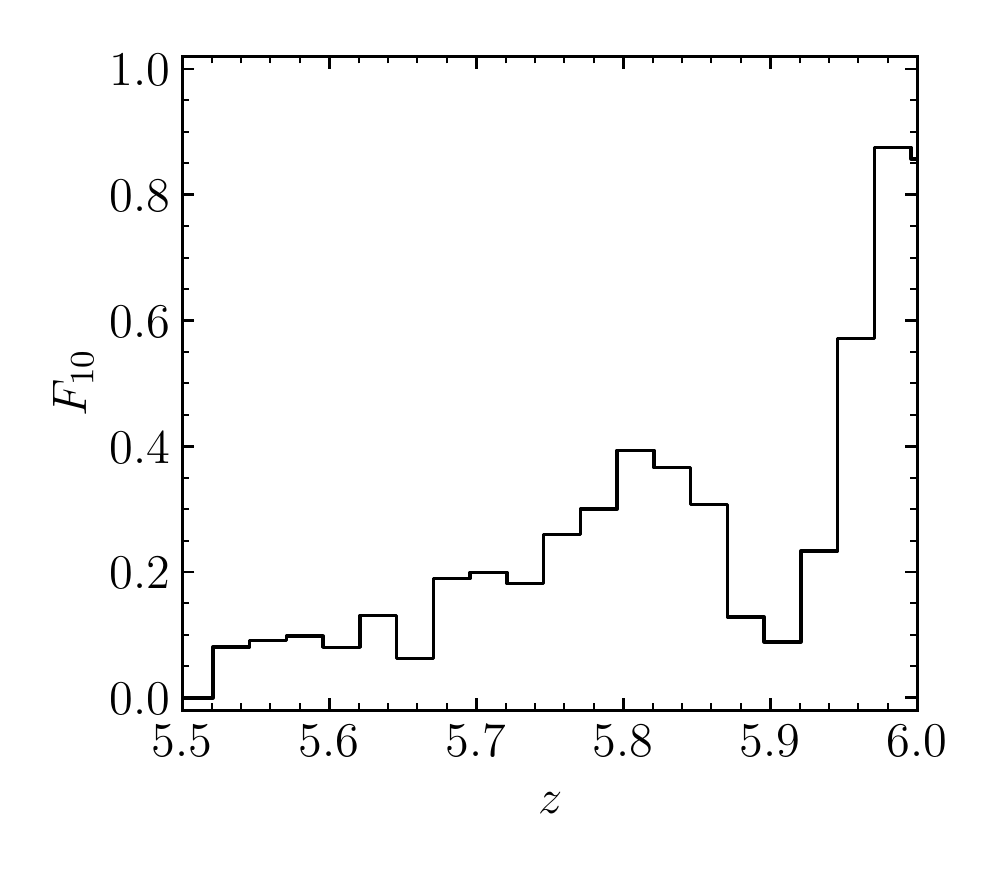}{2.6in}{(c)}
    }
    \caption{Observed \lyb\ dark gaps. 
    \textbf{{(a)}} Gap length versus central redshift. Dark gaps cut at the red end by the proximity zone cut are labeled with arrows.  
    \textbf{{(b)}} Cumulative distribution of dark gap length for two redshift intervals. The upper/lower bounds of the shaded region are described in Section~\ref{sec:gap_statistics}.
    \textbf{{(c)}} The fraction of QSO spectra showing dark gaps longer than $10 \cmpch$ as a function of redshift, $F_{10}$. $F_{10}$ is plotted with a binning of $\Delta z = 0.025$. See text and Appendix \ref{app:F01} for discussions about the drop of $F_{10}$ near $z=5.9$.
    \\
    (The data used to create this figure are available online at \url{https://ydzhuastro.github.io/lyb.html}.)
    \label{fig:obs}} 
\end{figure*}

Figure~\ref{fig:obs} displays the statistical properties of dark gaps detected in the \lyb\ forest from our sample. We detect 195 dark gaps in total, of which 24 have $L \geq 10 \cmpch$. Panel (a) plots length versus central redshift of these dark gaps. Long dark gaps become less common as redshift decreases. Nevertheless, some long gaps still exist down to $z<5.6$.

We calculate the cumulative distribution function of dark gap length, $P(<L)$. Dark gaps are sorted into two redshift bins according to their redshifts at both ends. For distributions that include dark gaps cut at the red end by the proximity zone limit, we calculate a lower bound on $P(<L)$ by assuming a infinite length for these gaps, and an upper bound by assuming the gap length that would appear in the absence of any proximity effect is the same as the one measured. As shown in Figure~\ref{fig:obs} (b), longer dark gaps become more numerous over $5.75 < z < 6.00$ compared to $5.50 < z < 5.75$. This significant evolution of $P(<L)$ is consistent with the results shown in panel (a).

Following \citetalias{zhu_chasing_2021}, we quantify the spatial coverage of large \lyb-opaque regions by calculating the fraction of QSO spectra showing long ($L\geq10\cmpch$) dark gaps as a function of redshift, $F_{10}(z)$. Here we use $10\cmpch$ as the threshold because dark gaps longer than this in the late reionization models (see Section~\ref{sec:models}) are dominated by those containing neutral islands. Based on our tests, the number of dark gaps predicted by different models differs the most for $L\gtrsim10\cmpch$, as also suggested by \citet{nasir_observing_2020}.

We calculate $F_{10}$ \add{at each redshift} and average over $\Delta z=0.025$ bins. As shown in Figure~\ref{fig:obs} (c), $F_{10}$ has a significant redshift evolution over $5.5<z<6.0$. It increases rapidly with redshift over $5.5<z<5.8$, from $\sim 10\%$ to $\sim 40\%$, and climbs up to $\sim80\%$ by $z=6.0$ after a drop at $z\sim5.9$. The reason of the drop is unclear,
but the limited number of QSO sightlines may produce large statistical fluctuations (Appendix~\ref{app:F01}), as also shown in the model predictions in Section~\ref{sec:models}. For comparison, we compute $F_{01}$, the fraction of QSO sightlines exhibiting dark gaps with $L\geq 1\cmpch$, and find no significant drop near $z=5.9$ (see Appendix~\ref{app:F01}).

\subsubsection{Long Dark Gap toward PSO J025\texorpdfstring{$-$}{Lg}11}
We find a particularly interesting \lyb\ gap toward the $z=5.849$ QSO PSO J025$-$11 (Figure~\ref{fig:spectra}). This dark gap is within the redshift interval of a long ($68\cmpch$) trough in the \lya\ forest. It spans $5.526\leq z \leq 5.613$ with a length of $28\cmpch$. This is longer and extending to a even lower redshift than the 19 and $23\cmpch$ \lyb\ troughs with $z_{\rm min}=5.66$ within the extreme ($110\cmpch$) \lya\ trough over $5.523 \leq z \leq 5.879$ toward ULAS J0148$+$0600 \citep{becker_evidence_2015}.  
This dark gap toward J025$-$11 contains no apparent transmission peaks, even in the unbinned data. The $2\sigma$ lower limit of $\teff \geq 6.067$ measured over the complete trough indicates that it is highly opaque. 

There is a possibility that part of the trough may be due to either \lyb\ or foreground \lya\ absorption from the circum-galacitc medium (CGM) around intervening galaxies. In this case, corresponding metal lines may be present. We check for potential CGM absorption using the XQR-30 metal absorber catalog (Davies et al., in prep; see Appendix \ref{app:moreMetal} for technique details). We find no intervening metal systems within the redshift range of the gap. We note that this line of sight has a DLA near the redshift of the QSO, as evidenced by the damping wing at the blue edge of the \lya\ proximity zone.  The \lyb\ gap described here is at a velocity separation of $>$3000$~{\rm km\,s^{-1}}$ from the QSO, however, and is unaffected by the DLA. The XQR-30 catalog does include a \ion{C}{4} system towards J025$-$11 at $z=4.5138$, for which \lya\ would fall at the blue end of the \lyb\ trough.
Overall, however, the general lack of metal absorbers associated with this long dark gap may suggest that the gap probes a low-density region. This would favor the association of highly opaque sightlines with galaxy underdensities, as seen in imaging surveys for galaxies surrounding long \lya\ troughs  \citep[][]{becker_evidence_2018,kashino_evidence_2020,christenson_constraints_2021}. 

We examine the possible role of metal-enriched absorbers more broadly in Appendix~\ref{app:moreMetal}, finding little evidence for a strong correlation with long \lyb\ troughs.  We also examined a sample of lower-redshift lines of sight in Appendix~\ref{app:lowz}, finding that metal-enriched absorbers in the foreground \lya\ alone are unlikely to create such a long dark gap.  We emphasize that this gap falls in redshift within a long \lya\ trough spanning $5.461\leq z \leq5.674$ with $L=68\cmpch$ that does not appear to be affected by metal absorbers (\citetalias{zhu_chasing_2021}).  This combination of factors gives us confidence that the $L=28\cmpch$ dark gap is genuinely caused by IGM opacity \footnote{In an extreme case where this foreground absorber links two shorter dark gaps, although very unlikely, one of these two shorter dark gaps would still have a size of $L\sim25\cmpch$.}.

\section{Models and Discussion\label{sec:models}}
\subsection{Models and Mock Spectra \label{sec:mock_spectra}}
Here we compare our results to predictions from models based on hydrodynamical simulations. 
We use the following models, which were also used in \citetalias{zhu_chasing_2021}: 
\begin{enumerate}
\item the {\tt homogeneous-UVB} model \add{from the Sherwood Simulation Suite} \citep[][]{bolton_sherwood_2017}\add{, which uses a homogeneous \citet{haardt_radiative_2012} UV background}; 
\item late reionization models wherein reionization ends at $z \lesssim 5.3$, including {\tt K20-low-{$\tau_{\rm CMB}$}}, {\tt K20-low-{$\tau_{\rm CMB}$}-hot}, {\tt K20-high-{$\tau_{\rm CMB}$}} models from \citet{keating_constraining_2020}, and {\tt ND20-late-longmfp}, {\tt ND20-late-shortmfp} models from \citet{nasir_observing_2020}; and 
\item an early reionization model wherein the IGM is mostly ionized by $z=6$ but large scale fluctuations in the UVB, which are amplified by a short mean free path of ionizing photons ($\lambda_{\rm mfp}^{912}=10\cmpch$ at $z=5.6$), persist down to lower redshifts ({\tt ND20-early-shortmfp}, \citealp{nasir_observing_2020}).   
\end{enumerate}
\add{These models were chosen, in part, because they reproduce at least some other observations of the \lya\ forest. The {\tt homogeneous-UVB} model agrees well with observations at $z<5$ including the IGM temperature and flux power spectra \citep{bolton_sherwood_2017}, although it fails to predict the \lya\ opacity distribution at $z>5$ \citep[e.g.,][]{bosman_hydrogen_2021}. The late reionization and fluctuating UVB models are broadly consistent with observations of IGM temperature, mean \lya\ transmission, and fluctuations in \lya\ opacity over the redshift range we are interested in \citep{keating_constraining_2020,nasir_observing_2020}. Moreover, these models are able to produce long \lya\ troughs at $z<6$ (e.g., \citetalias{zhu_chasing_2021}). We note that, nevertheless, that none of the models we use can self-consistently predict the mean free path evolution over $5<z<6$ as measured in \citet{becker_mean_2021}.}

\add{In the {\tt homogeneous-UVB} model, the IGM is instantaneously reionized at $z=15$. At $z<6$, therefore, the IGM in this model is fully ionized and the gas is hydrodynamically relaxed.  A homogeneous UVB model that produced a later reionization \citep[e.g.,][]{puchwein_consistent_2019,villasenor_inferring_2021} would mainly alter the temperature and pressure smoothing at $z < 6$.  These are small-scale effects, however, that should only minimally impact our measurements.  We would generally expect that any homogeneous UVB model that reionizes by $z = 6$ would produce similar dark gap statistics as the \citet[][]{haardt_radiative_2012} UVB once the ionization rates at $z < 6$ are rescaled to reproduce the observed mean flux.}

The {\tt K20-low-{$\tau_{\rm CMB}$}-hot} model shares a similar reionization history with the {\tt K20-low-{$\tau_{\rm CMB}$}} model, but it has a different thermal history with a volume-weighted mean temperature at the mean density at $z=6$ of $T_0=10,000$ K compared to that of the latter being 7000 K. The {\tt K20-high-{$\tau_{\rm CMB}$}} model has an earlier mid-point of reionization at $z=8.4$, which is at the upper end of the value suggested by CMB measurements \citep{planck_collaboration_planck_2020}. As for the late reionization models from \citet{nasir_observing_2020}, the main difference is that the {\tt ND20-late-shortmfp} model includes stronger post-reionization UVB fluctuations than the {\tt ND20-late-longmfp} model, while they both have neutral islands surviving at $z<6$. The mean free path of ionizing photons at $z=5.6$ in these two models are $\lambda_{\rm mfp}^{912}=10$ and 30 $\cmpch$, respectively. 

The box sizes we use in this work are $L=160$, 160, and 200 $\cmpch$, for simulations in \citet[][]{bolton_sherwood_2017}, \citet{keating_constraining_2020}, and \citet{nasir_observing_2020}, respectively. We note that the {\tt K20} models are from radiative transfer simulations run 
in post-processing and that the {\tt ND20} models are semi-numeric models. For more details on the models, see \citetalias{zhu_chasing_2021}.

\add{We rescale the optical depths in the simulations as needed in order to match the observed mean flux in the \lya\ forest \citep[see, e.g., \S2.2 in][and references therein]{bolton_sherwood_2017}.  We scale to the measurements of \cite{bosman_new_2018}, which are consistent with the mean \lya\ fluxes obtained from our sample.  The same rescaling factor is then applied to both the \lya\ and corresponding \lyb\ optical depths.  We note that this rescaling mainly applies to the {\tt homogeneous-UVB} model, for which scaling by factors of $\sim$0.4$-$0.6 is required over $5<z<6$.  We are therefore explicitly testing only a {\tt homogeneous-UVB} model that also matches the observed mean \lya\ flux.}  The \citet{keating_constraining_2020} models already produce a mean \lya\ transmission consistent with the measurements of \citet{bosman_new_2018}.  The mock spectra from this simulation are continuous in redshift, with a smoothly evolving mean flux.   \add{\citet[][]{nasir_observing_2020} also calibrated their simulations to the observed $\teff$ from \citet[][]{bosman_new_2018} but provide one-dimensional skewers at discrete redshifts. For these simulations we therefore only need to rescale the optical depths such that the mock spectra described below have a mean flux that evolves smoothly with redshift.}

We derive dark gap predictions from forward-modeled mock spectra that are matched to the observed sample in QSO redshift, resolution, and S/N.  \citet{keating_constraining_2020} provide mock spectra of the \lyb\ forest including the foreground \lya\ absorption.  \add{For the {\tt homogeneous-UVB} model and models from \citet{nasir_observing_2020}} we follow the methods described in \citetalias{zhu_chasing_2021} to build the mock \lyb\ forest and foreground \lya\ forest.  In all cases we re-bin the mock spectra and apply the noise arrays according to each observed spectrum. 


\subsection{Model Comparisons}
\subsubsection{Comparisons of \texorpdfstring{$F_{10}$}{Lg}}

\begin{figure*}[htb!]
    \centering \includegraphics[width=6.6in]{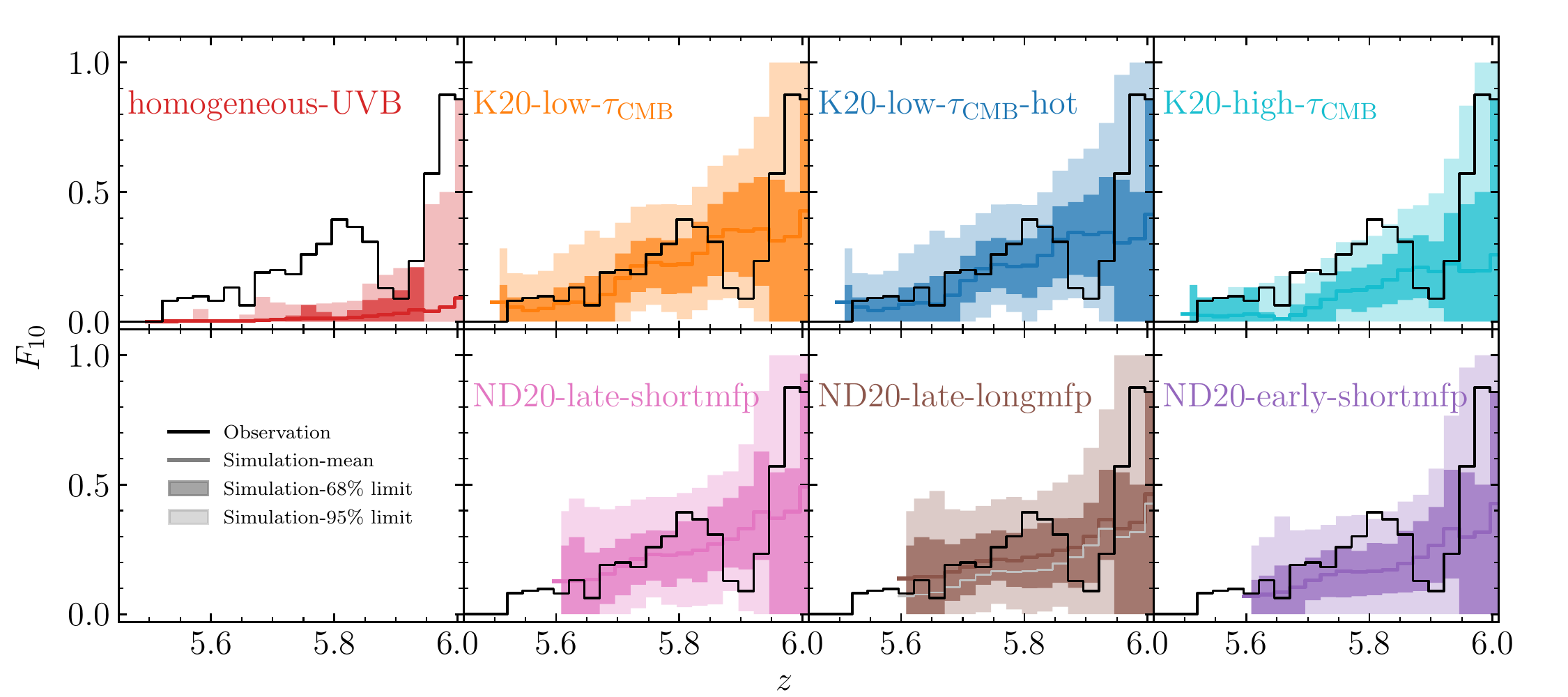}
    \caption{Fraction of QSO spectra showing long ($L\geq 10\cmpch$) \lyb\ dark gaps as a function of redshift predicted by different models. The colored lines, dark-shaded regions, and light-shaded regions show the mean, 68\%, and 95\% limits of $F_{10}$ predicted by models based on 10,000 realizations. The black lines plot $F_{10}$ from the observations. For comparison, we also overplot the mean $F_{10}$ predicted by the {\tt ND20-early-shortmfp} in the panel of the {\tt ND20-late-longmfp} model with a gray line.
    }
    \label{fig:F10}
\end{figure*}

We compute the predicted $F_{10}$ of each model based on 10,000 randomly selected sets of mock spectra of the same size as the observed sample.  Their mean, 68\%, and 95 \% limits are plotted in Figure~\ref{fig:F10}, along with the observations. Similar to $F_{30}$ 
\footnote{$F_{30}$ is defined as the fraction of QSO spectra exhibiting gaps longer than $30\cmpch$ as a function of redshift.}
for the \lya\ forest in \citetalias{zhu_chasing_2021}, $F_{10}$ shows jagged features due to the combined effects of step changes in the number of sightlines with redshift and the quantization of $F_{10}$ for a finite sample size. We note that 68 and 95 percentile limits can share their upper and/or lower bounds at some redshifts, for the same reason. These features, on the other hand, show the constraining ability of the current sample size and data quality. The drop at $z \sim 5.9$ seen in the observed $F_{10}$ is also broadly included in the 95\% limits for most of the models. 

The {\tt homogeneous-UVB} model is not supported by the data at the $\geq95\%$ level. 
This is consistent with the conclusion based on the \lya\ forest in \citetalias{zhu_chasing_2021} that a fully ionized IGM with a homogeneous UVB scenario is disfavored by the data at $z<6$ down to $z\simeq5.3$. In contrast, the late reionization models are still consistent with the data, except for the {\tt K20-high-$\tau_{\rm CMB}$} model, which covers the observed $F_{10}$ just within its 95\% upper limit. This supports the conclusion of \citetalias{zhu_chasing_2021} that this extended reionization model is less favored by the data due to its insufficient neutral hydrogen and/or UVB fluctuations at $z<6$.

Our results further show that dark gaps in \lyb\ are more sensitive probes of neutral regions than gaps in \lya.
For dark gaps in the \lya\ forest, we see little difference between the {\tt ND20-early-shortmfp} model and the {\tt ND20-late} models (\citetalias{zhu_chasing_2021}). In the \lyb\ forest, however, the former predicts a smaller $F_{10}$ than the latter by $\sim 0.05$ at most redshifts. This difference is not enough for us to distinguish them based on the current sample, although the \lyb\ gaps put some pressure on the early reionization model. \citet{nasir_observing_2020} note that these models are different in their \lyb\ dark gap length distributions while they cannot be distinguished in the \lya\ forest. Nevertheless, we compute the differential dark gap length distribution for individual $\Delta L$ bins, $L\Delta P(L)/\Delta L$, and find their differences are minor compared to the scatter of the data. Looking ahead to the era of ELTs, we forecast that $\sim 100$ lines of sight with the \lyb\ forest covering $z\sim 5.8$ would be needed to distinguish the two models at $\sim$95\% confidence based on $F_{10}$. A signal-to-noise ratio of 50 per $10\,\rm km\,s^{-1}$ for the spectra would be adequate according to our tests using mock spectra.

\subsubsection{Total Number of Long Dark Gaps at \texorpdfstring{$z \leq 5.8$}{Lg}}
\begin{figure*}[htb!]
    \begin{center}
    \gridline{
    \fig{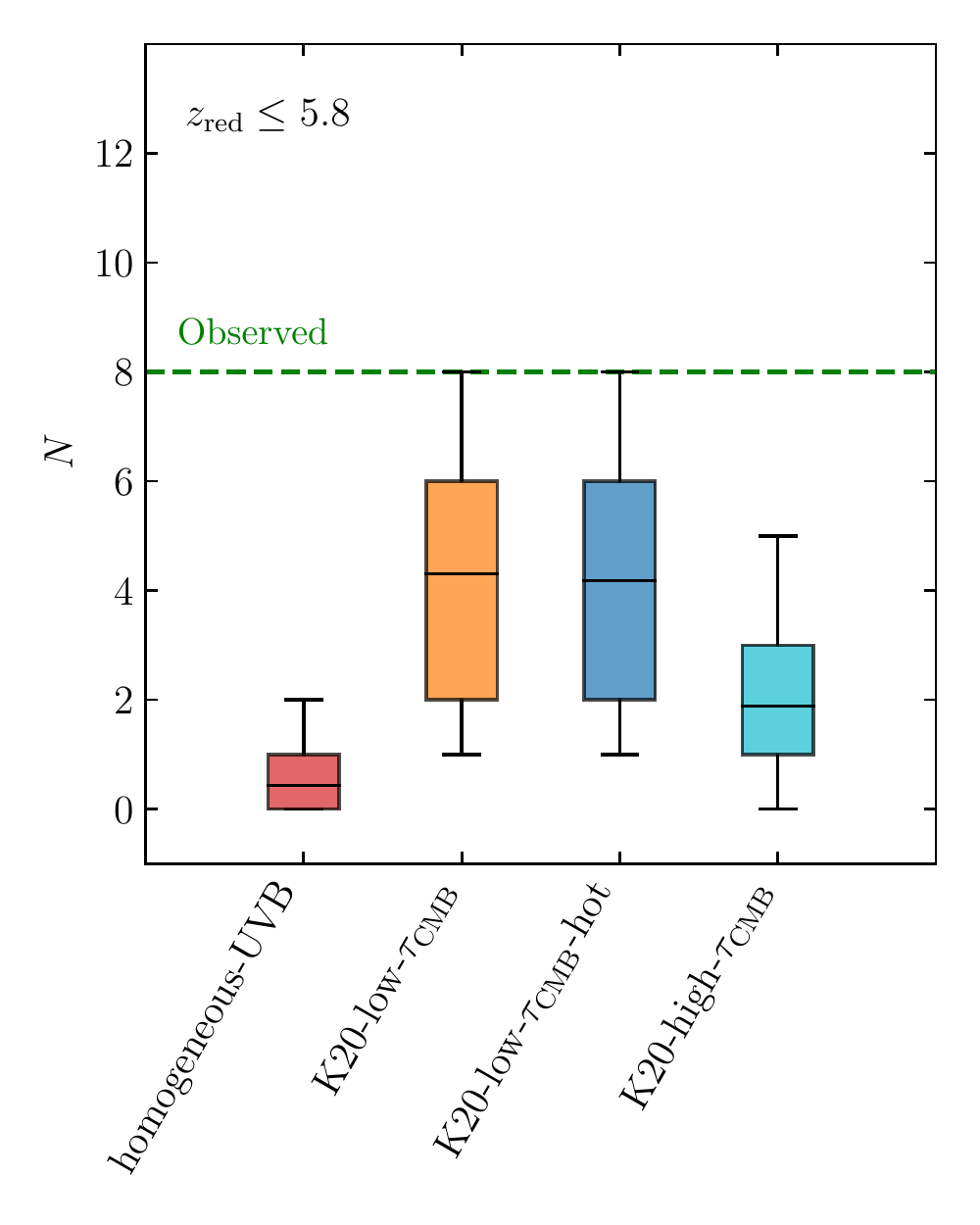}{2.6in}{(a)}
    \fig{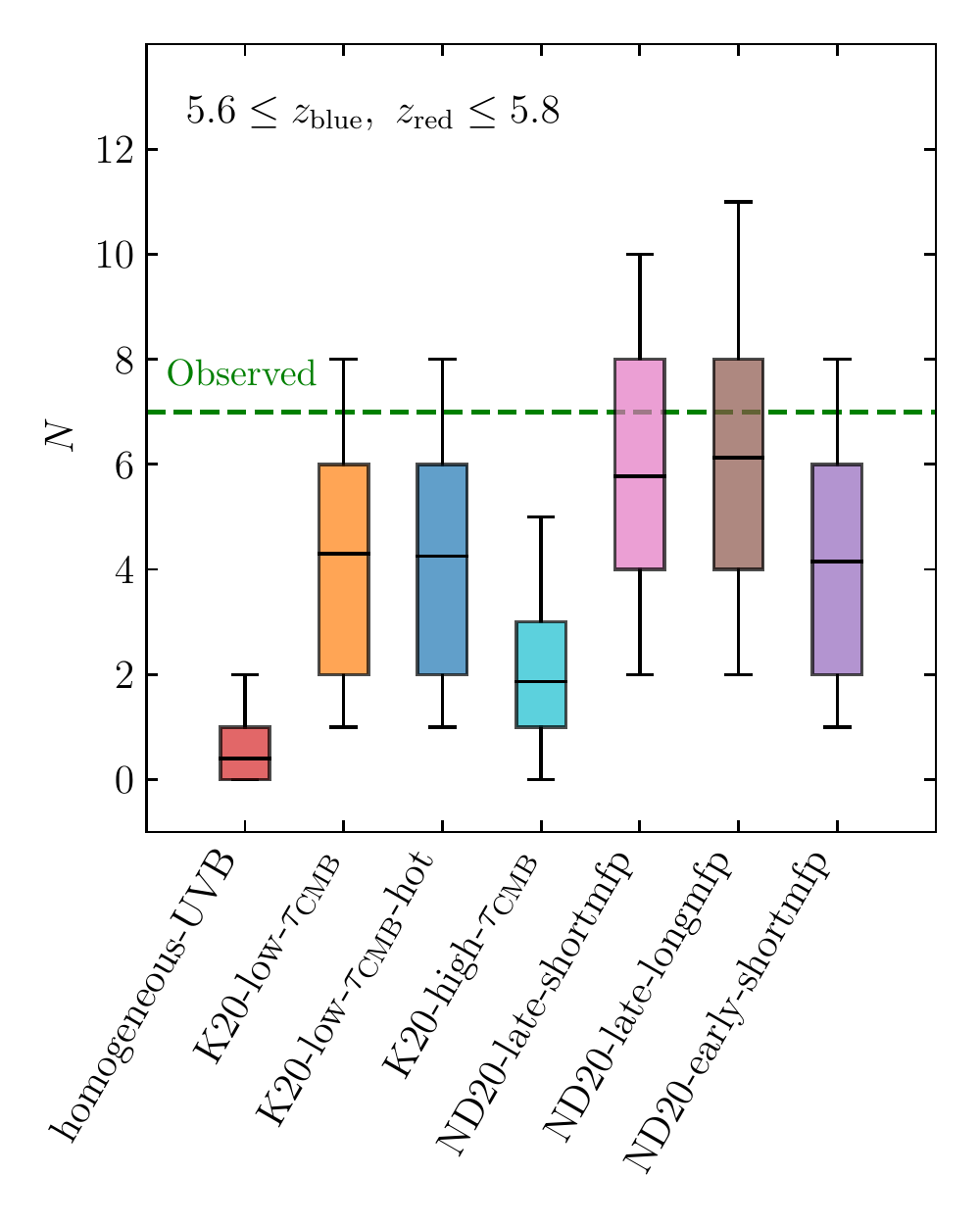}{2.6in}{(b)}
    }
        \caption{
        \textbf{(a) }
        Number of long ($L\ge10\cmpch$) \lyb\ dark gaps that lie entirely below $z=5.8$ in the mock sample from each model.
        \textbf{(b) }
        Number of long dark gaps entirely over $5.6\le z \le 5.8$. 
        In both panels the center lines, boxes, and error bars show the mean, 68\% limits, and 95\% limits, respectively. The observed numbers of long dark gaps in each redshift range are indicated by dashed green lines.
        \label{fig:ccount_10_zlt58}}
    \end{center}
\end{figure*}
To further illustrate the differences between models, we use our mock data to calculate the total number of long dark gaps predicted to lie entirely at $z < 5.8$. Figure~\ref{fig:ccount_10_zlt58} compares the model results to the observations. Given that the {\tt ND20} models only have outputs down to $z=5.6$, we exclude these models when counting the total number of long dark gaps below $z = 5.8$. We nevertheless include the {\tt ND20} models for dark gaps that fall entirely over $5.6 \le z \le 5.8$ for reference.

The results are consistent with those from $F_{10}$. As shown in Figure~\ref{fig:ccount_10_zlt58} (a), the 95\% upper limit of the predicted number of long dark gaps by the {\tt homogeneous-UVB} model is 2. This is a factor of 4 smaller than the observed value, which is 8.  The {\tt K20-high-$\tau_{\rm CMB}$} model is also disfavored by the data at $>95\%$ confidence given its deficit of long dark gaps. 
In contrast, the rapid late reionization models, i.e. {\tt K20-low-$\tau_{\rm CMB}$} models, agree with the observations within their $95\%$ limits. 

Over $5.6 \le z \le 5.8$ the observed number of long dark gaps decreases by one while the simulation predictions have little change. In this case, rapid late reionization models from \citet[][]{keating_constraining_2020} are still consistent with the data. The observations also support both fluctuating UVB and late reionization models from \citet[][]{nasir_observing_2020}. We note that the difference between the predicted mean numbers and the observed value is smallest for the {\tt ND20-late} models, wherein $ \langle x_{\rm HI} \rangle$ is still higher than $\sim5\%$ by $z=5.6$. On the other hand, the {\tt K20-high-$\tau_{\rm CMB}$} model is disfavored by the data also in this redshift range. 
This would suggest that very extended reionization scenarios in which insufficient neutral hydrogen and/or UVB fluctuations remain at $z<6$ may be disfavored.

\subsubsection{Detection Rate of an \texorpdfstring{$L\geq 28\cmpch$}{Lg} Dark Gap \label{sec:detection-rate}}

\begin{figure}[htb!]
\begin{center}
    \includegraphics[width=0.45\textwidth]{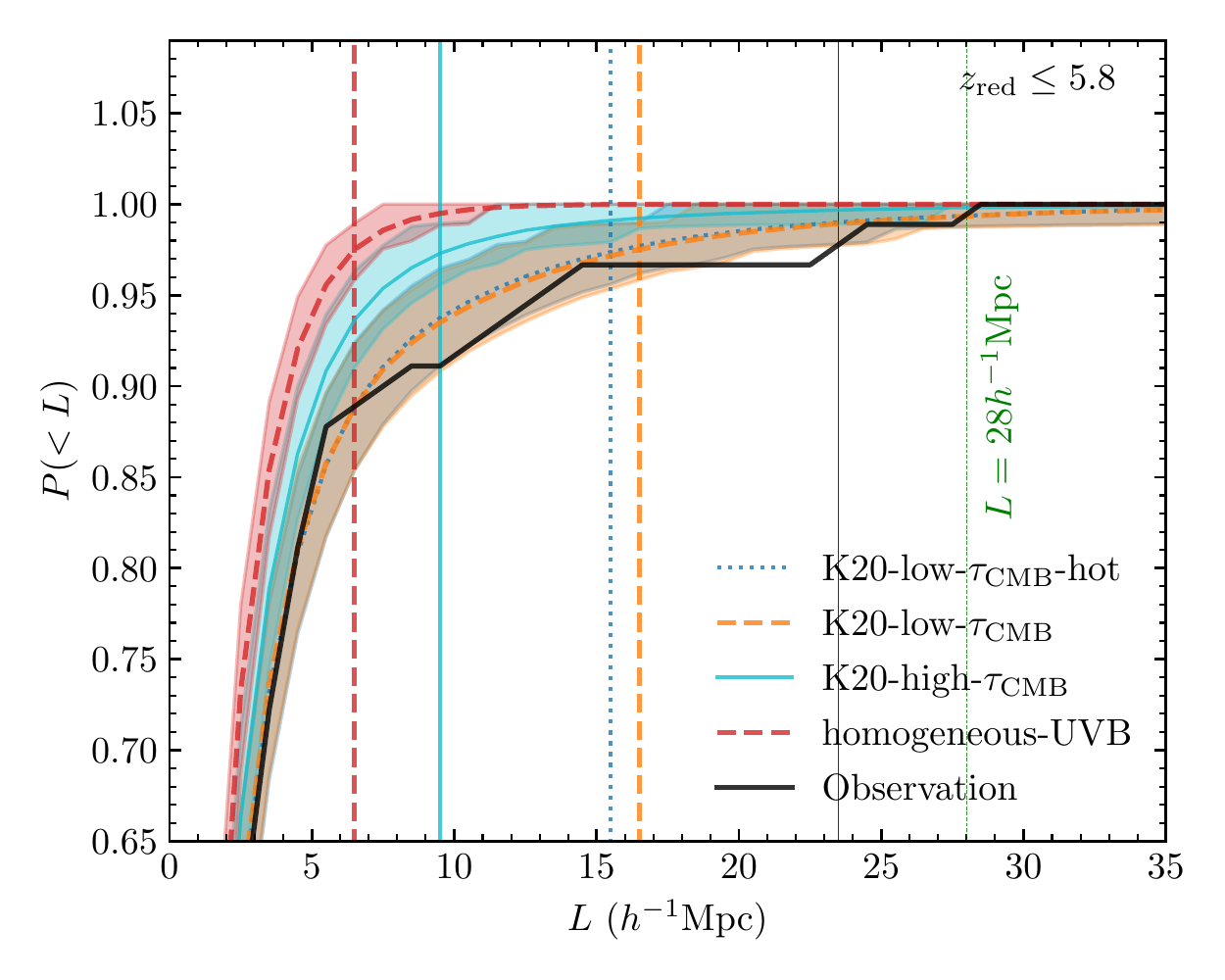}
    \caption{
    Cumulative distribution of dark gaps that are entirely below $z=5.8$. Vertical colored lines indicate the 97.5 percentile limit for each model.
    The color shaded regions plot the 68\% limit for each model. From left to right are the {\tt homogeneous-UVB} model, the {\tt K20-high-$\tau_{\rm CMB}$} model, and the {\tt K20-low-$\tau_{\rm CMB}$-hot} model (almost completely overlapped with the {\tt K20-low-$\tau_{\rm CMB}$} model), respectively.
    \label{fig:detectionrate}}
    \end{center}
\end{figure}
Perhaps the most conspicuous feature in the observations is the $L=28\cmpch$ dark gap toward PSO J025$-$11 that extends down to $z \simeq 5.5$.  The appearance of this gap may indicate that significant neutral hydrogen islands and/or UVB fluctuations persist down to $z \simeq 5.5$, and provide further leverage for discriminating between models. As the outputs of the {\tt ND20} models have no redshift coverage for this dark gap, we only compare the {\tt K20} models and the {\tt homogeneous-UVB} model for this section. 

For each model, we use 10,000 bootstrapping realizations to calculate the cumulative distribution function (CDF) of dark gap length, $P(<L)$. Figure~\ref{fig:detectionrate} compares the observed and predicted $P(<L)$ for dark gaps that are entirely below $z=5.8$. As indicated by the vertical lines, the observed dark gap with $L=28\cmpch$ is well beyond the $95\%$ limits of all the models shown here. These results suggest that the $L=28\cmpch$ gap we observed in the \lyb\ forest toward PSO J025$-$11 is extremely rare in these models. We perform Mann-Whitey U tests \citep[][]{mann_test_1947} for the hypotheses that the distributions of $L$ in the data and predicted by models are equal, for each model respectively. 
The hypothesis is rejected with $p$-values $<0.05$ for the {\tt homogeneous-UVB} model. 

We further calculate the detection rate of at least one $L=28\cmpch$ gap entirely below $z=5.61$ in the mock samples from each model with the required redshift coverage. We note that in the data there are 10 QSO spectra where the \lyb\ forest (excluding the proximity zone) covers the full central redshift range of the $L=28\cmpch$ dark gap. We find zero detections of such long dark gaps in the {\tt homogeneous-UVB} model out of 10,000 trails. The {\tt K20-high-$\tau_{\rm CMB}$} model yields a detection rate of $4\%$. Both the {\tt K20-low-$\tau_{\rm CMB}$} and {\tt K20-low-$\tau_{\rm CMB}$-hot} models give higher detection rate of $10\%$. These results suggest that in the case of a late reionization, models with a volume-weighted average neutral hydrogen fraction $\langle x_{\rm HI}\rangle \gtrsim 5\% $ at $z=5.6$ are more consistent with the observations. In addition, the relatively rare presence of $L\geq 28 \cmpch$ gaps in the models could also be related to the size of the simulation volume ($160 \cmpch$ for the {\tt K20} simulations). Simulations run in larger volumes may be needed to compute more accurate statistics on the incidence of these rare, long troughs in late reionization models. %

\subsection{Neutral Hydrogen Fraction \label{sec:fHI}}

\begin{figure*}
    \centering
    \gridline{
        \fig{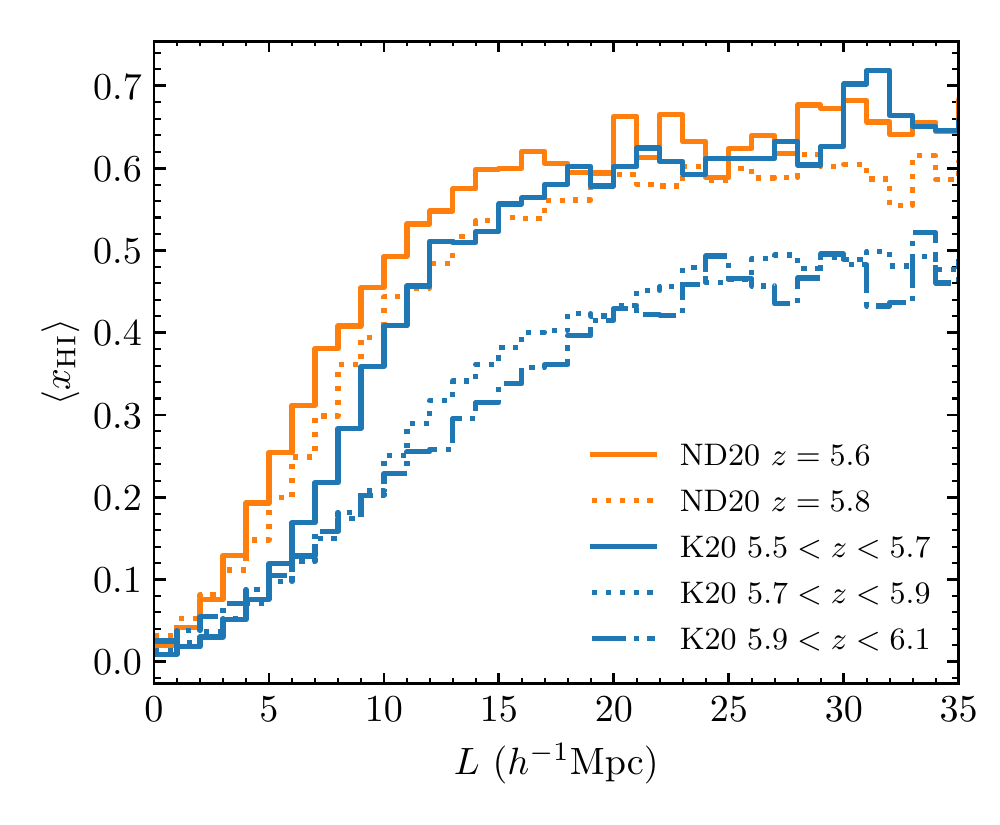}{3in}{(a)}
        \fig{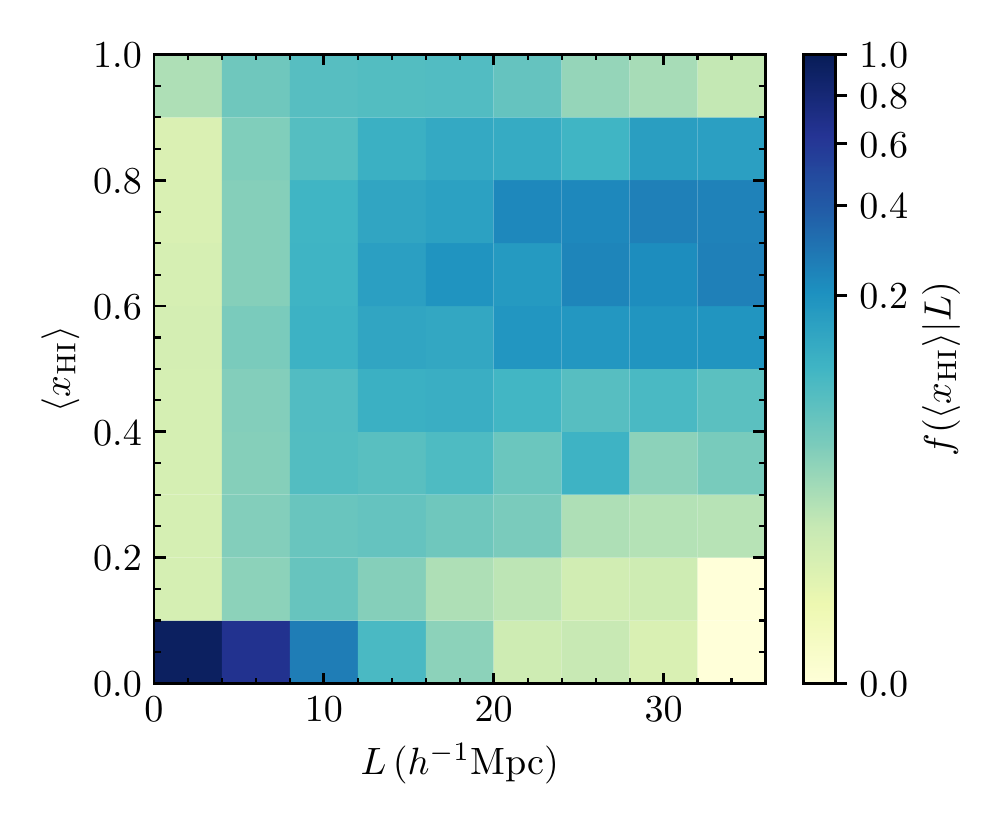}{3.1in}{(b)}
    }
    \caption{(a) The mean volume-weighted neutral fraction ($ \langle x_{\rm HI} \rangle $) over a \lyb\ dark gap for a given dark gap length in the mock data. In this figure, ``ND20'' and ``K20'' refer to the {\tt ND20-late-longmfp} model and the {\tt K20-low-$\tau_{\rm CMB}$} model, respectively.
    (b) Distribution of $ \langle x_{\rm HI} \rangle $ for a given \lyb\ dark gap length, $f(x_{\rm HI}  | L)$, in the {\tt ND20-late-longmfp} model at $z=5.6$. $f(x_{\rm HI}  | L)$ is normalized for each $L$ interval.
    }
    \label{fig:xHIvsL}
\end{figure*}

\begin{figure}
    \centering
    \includegraphics[width=3.3in]{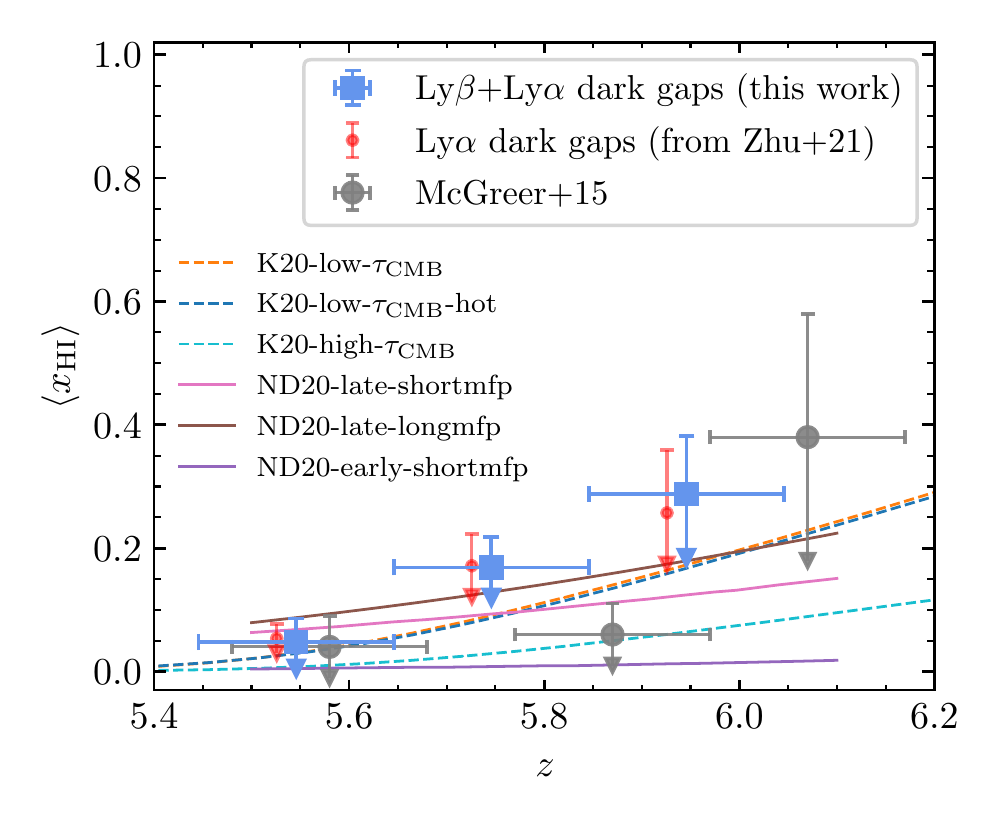}
    \caption{Inference on the neutral hydrogen fraction ($\langle x_{\rm HI} \rangle$) from \lyb\ dark gaps, which are also dark in the \lya\ forest by definition. We show constraints based on \lya\ dark gaps from \citetalias{zhu_chasing_2021} with red error bars, which are shifted by $-0.02$ in redshift for display. Gray markers plot the constraints based of fraction of pixels that are dark in both the \lya\ and \lyb\ forests from \citet[][]{mcgreer_model-independent_2015}. The vertical error bars show the $68\%$ ($1\sigma$) limits. The horizontal error bars indicate the $\Delta z=0.2$ redshift bins. \add{For reference, colored lines plot the redshift evolution of $\langle x_{\rm HI} \rangle$ for the reionization models used in this work.}}
    \label{fig:fHI}
\end{figure}

We can further use dark gaps to infer an upper limit on $\langle x_{\rm HI} \rangle$.  One can set a strict upper limit on the neutral fraction by assuming that all dark gaps correspond to neutral gas \citep[e.g.,][]{mcgreer_first_2011,mcgreer_model-independent_2015}.  At the end of reionization, however, a combination of density and UVB fluctuations will tend to produce dark gaps even once the gas is ionized.  We therefore wish to use insights from reionization models to derive a more physically motivated but still conservative upper limit on $\langle x_{\rm HI} \rangle$ from the covering fraction of dark gaps.  As described below, we use the fact that dark gaps in the late reionization models tend to show a correlation between the volume-averaged neutral fraction within a gap and the gap length.  By applying this relationship to the observed gap length distribution we can set constraints on $\langle x_{\rm HI} \rangle$.

Our goal is to set physically reasonable constraints on $\langle x_{\rm HI} \rangle$ while minimizing the model dependency. We therefore wish to identify the maximum average neutral fraction for a given gap length that is allowed by the models.  We first explore the distribution of neutral fractions for a given dark gap length, $f(x_{\rm HI}  | L)$. %
We focus on two models wherein neutral regions contribute significantly to forming dark gaps, the {\tt ND20-late-longmfp} model and the {\tt K20-low-$\tau_{\rm CMB}$} model. Using the mock data, we calculate $x_{\rm HI}$ for each dark gap by averaging the neutral fraction pixel-wise. Figure~\ref{fig:xHIvsL} (a) plots the mean neutral fraction of dark gaps as a function of length, $ \langle x_{\rm HI}\rangle _L $, at different redshifts.  It is related to $f(x_{\rm HI}  | L)$ by
\begin{equation}
    \langle x_{\rm HI} \rangle _L = \int_0^1 x_{\rm HI} f(x_{\rm HI} | L)  d x_{\rm HI}.
    \label{eq:<xHI>}
\end{equation}
As shown in the figure, dark gaps of a given length tend to be more neutral as redshift decreases. 
This is largely because the opacity of the ionized IGM tends to decrease with decreasing redshift, making it more difficult to produce long gaps through density and/or UVB fluctuations alone. 
In order to set conservative upper limits of $\langle x_{\rm HI} \rangle$ we adopt the $\langle x_{\rm HI} \rangle_L$ relationship from {\tt ND20-late-longmfp} at $z = 5.6$. The normalized $f(x_{\rm HI}  | L)$ for each dark gap length interval is plotted in Figure~\ref{fig:xHIvsL} (b).  This is similar to but slightly higher than the relationship from {\tt K20-low-$\tau_{\rm CMB}$} at the same redshift.  We also note that the redshift evolution in $\langle x_{\rm HI} \rangle_L$ in these models is relatively modest, up to a factor of $\sim$2 in the {\tt K20-low-$\tau_{\rm CMB}$} model between $z \sim 6$ and 5.6.

In order to translate the observed gap length distribution into a $\langle x_{\rm HI} \rangle$ constraint, we calculate $\mathscr{F}_L$, the fraction of QSO spectra showing dark gaps with length $L$ as a function of redshift. At a certain redshift, the total mean neutral hydrogen fraction is then given by
\begin{equation}
    \langle x_{\rm HI} \rangle = \sum_{L=1}^{\infty} \mathscr{F}_L \langle x_{\rm HI} \rangle _L.
    \label{eq:fHI}
\end{equation}
Here we use a sum for $L$ instead of an integral because we measure dark gap lengths in increments of $1\cmpch$. 
To estimate the uncertainty in $\langle x_{\rm HI} \rangle$, we randomly select the observed sightlines with replacement and calculate the corresponding $\mathscr{F}_L$. We use bootstrapping to randomly sample the neutral hydrogen fraction from $f( x_{\rm HI} | L)$ given by models and multiply by the observed $\mathscr{F}_L$ of this sample, then sum up for all dark gap lengths. The final uncertainty in $\langle x_{\rm HI} \rangle$ is calculated by repeating this process 10,000 times.

The results are shown in Figure~\ref{fig:fHI}. We calculate $\langle x_{\rm HI} \rangle$ in \add{Equation} (\ref{eq:fHI}) over $\Delta z=0.2$ bins. The inferred upper limits on $\langle x_{\rm HI} \rangle$ are $0.05_{-0.04}^{+0.04}$, $0.17_{-0.05}^{+0.05}$, and $0.29_{-0.10}^{+0.09}$ at $z\simeq 5.55$, 5.75, and 5.95, respectively. We also calculate $\langle x_{\rm HI} \rangle$ following the same method based on \lya\ dark gaps presented in \citetalias{zhu_chasing_2021},  as shown with red symbols in Figure~\ref{fig:fHI}. 
The \lya\ dark gaps yield $\langle x_{\rm HI} \rangle \leq$ $0.05$, $0.17$, and $0.26$ at $z\simeq 5.55$, 5.75, and 5.95, respectively. 
The measurements based on \lya\ and \lyb\ dark gaps are highly consistent with each other. Compared to the measurements based the fraction of dark pixels by \citet{mcgreer_model-independent_2015}, our results potentially allow a higher neutral fraction over $5.6 \lesssim z \lesssim 6.0$ and a later reionization. The difference in $\langle x_{\rm HI} \rangle$ might be due to cosmic variance and/or the different definitions of dark gaps and dark pixels used in these works. The $\langle x_{\rm HI} \rangle$ measurement at $z\sim 5.9$ in \citet[][]{mcgreer_model-independent_2015}, moreover, may be biased by transmission peaks in the QSO proximity zone given that their wavelength range for both the \lya\ and \lyb\ forests ends at $z_{\rm QSO}-0.1$, which is less than 6.5 pMpc from the QSO at $z\sim 6$ (see proximity zone size measurements in e.g., \citealp{eilers_implications_2017,eilers_detecting_2020}). 

\section{Conclusion \label{sec:conslusion}}
In this work, we explore the IGM near the end of reionization using dark gaps in the \lyb\ forest over $5.5\lesssim z \lesssim 6.0$. We show that about 10\%, 40\%, and 80\% of QSO spectra exhibit long ($L\geq 10\cmpch$) dark gaps in their \lyb\ forest at $z\simeq5.6$, 5.8, and 6.0, respectively. Among these gaps, we detect a very long ($L=28\cmpch$) and dark ($\teff\gtrsim6$) \lyb\ gap extending down to $z\sim 5.5$ toward the $z_{\rm em}=5.85$ QSO PSO J025$-$11. 

A comparison between the observed \lyb\ dark gap statistics for the whole sample of 42 lines of sight and predictions from multiple reionization models \citep[][]{bolton_sherwood_2017,keating_constraining_2020,nasir_observing_2020} confirms that evidence of reionization in the form of neutral islands and/or a fluctuating UV background persists down to at least $z\sim 5.5$. This supports the conclusions in \citetalias{zhu_chasing_2021} and \citet{bosman_hydrogen_2021}.  In \citetalias{zhu_chasing_2021} we noted a possible tension between \lya\ gap statistics and a model wherein reionization ends by $z<6$ but has a relatively early mid-point of $z=8.4$ \citep{keating_constraining_2020}.  With \lyb\ this tension becomes more significant ($>95\%$ level) based on the count of long dark gaps at $z\le5.8$, suggesting that very extended reionization scenarios with insufficient remaining neutral hydrogen and/or UVB fluctuations at $z<6$ may be disfavored. In contrast, rapid late reionization models with $\langle x_{\rm HI} \rangle \gtrsim 5 \%$ at $z=5.6$ \citep[][]{keating_constraining_2020,nasir_observing_2020} are consistent with the observations. A model wherein reionization ends early but retains large-scale fluctuations in the ionizing UV background \citep[][]{nasir_observing_2020} is also permitted by the dark gap data. We note, however, that recent IGM temperature measurements from \citet{gaikwad_probing_2020} disfavor this model.

\add{
A caveat is that we are testing only specific reionization models, including only one with a fluctuating UVB in which reionization ends at $z > 6$.  By comparison, \citet{gnedin_cosmic_2017} showed that their full radiative transfer simulations, which reionized near $z \sim 7$, were able to reproduce the \lya\ dark gap distribution measured from ESI spectra of a set of twelve $z \sim 6$ QSOs. Because \lyb\ dark gaps are correlated with \lya\ opacities (Appendix \ref{app:Lyb_Lya}), it is possible that some early reionization scenarios with UVB fluctuations can reproduce our \lyb\ dark gap distributions while also matching the observed evolution of the mean \lya\ transmission. 
}

Finally, we use the observed \lyb\ gaps to place constraints on the neutral hydrogen fraction based on the association between neutral islands and dark gaps seen in reionization simulations. Our results are broadly consistent with, but more permissive than the constraints from \citet{mcgreer_first_2011,mcgreer_model-independent_2015} that are based on the dark pixel fraction.  Notably, we find an upper limit at $z\simeq5.75$ of $\langle x_{\rm HI} \rangle \leq 0.17$.  This constraint is consistent with scenarios wherein reionization extends significantly below $z = 6$.

\begin{acknowledgments}

    We thank Elisa Boera and Fahad Nasir for their help with simulated data and for useful discussions. \add{We also thank the anonymous referee for their constructive comments. }

    Y.Z., G.D.B., and H.M.C. were supported by the National Science Foundation through grants AST-1615814 and AST-1751404. H.M.C. was also supported by the National Science Foundation Graduate Research Fellowship Program under grant No. DGE-1326120.
    S.E.I.B. and F. Walter acknowledge funding from the European Research Council (ERC) under the European Union's Horizon 2020 research and innovation program (grant agreement No. 740246 ``Cosmic Gas'').
    L.C.K. was supported by the European Union's Horizon 2020 research and innovation program under the Marie Skłodowska-Curie grant agreement No. 885990.
    M.B. acknowledges support from PRIN MIUR project ``Black Hole winds and the Baryon Life Cycle of Galaxies: the stone-guest at the galaxy evolution supper'', contract $\#$2017PH3WAT.
    F.B. acknowledges support from the Australian Research Council through Discovery Projects (award DP190100252) and Chinese Academy of Sciences (CAS) through a China-Chile Joint Research Fund (CCJRF1809) administered by the CAS South America Center for Astronomy (CASSACA).
    H.C. thanks the support by NASA through the NASA FINESST grant NNH19ZDA005K.
    A.-C.E. acknowledges support by NASA through the NASA Hubble Fellowship grant $\#$HF2-51434 awarded by the Space Telescope Science Institute, which is operated by the Association of Universities for Research in Astronomy, Inc., for NASA, under contract NAS5-26555. 
    X.F. and J.Y. acknowledge support from the NSF grants AST 15-15115 and AST 19-08284. 
    M.G.H. acknowledges support from the UKRI STFC (grant Nos. ST/N000927/1 and ST/S000623/1).  
    G.K.'s research is partly supported by the Max Planck Society via a partner group grant.
    A.P. acknowledges support from the ERC Advanced Grant INTERSTELLAR H2020/740120.
    Parts of this work was supported by the Australian Research Council Centre of Excellence for All Sky Astrophysics in 3 Dimensions (ASTRO 3D), through project $\#$CE170100013.
    F. Wang thanks the support provided by NASA through the NASA Hubble Fellowship grant \#HST-HF2-51448.001-A awarded by the Space Telescope Science Institute, which is operated by the Association of Universities for Research in Astronomy, Incorporated, under NASA contract NAS5-26555.
    
    Based on observations collected at the European Southern Observatory under ESO programmes 060.A-9024(A), 084.A-0360(A), 084.A-0390(A), 084.A-0550(A), 085.A-0299(A), 086.A-0162(A),  
    086.A-0574(A), 087.A-0607(A), 088.A-0897(A), 091.C-0934(B), 096.A-0095(A), 096.A-0418(A), 
    097.B-1070(A), 098.A-0111(A), 098.B-0537(A), 0100.A-0243(A), 0100.A-0625(A), 0101.B-0272(A), 0102.A-0154(A), 0102.A-0478(A), 1103.A-0817(A), and 1103.A-0817(B).
    
    Some of the data presented herein were obtained at the W. M. Keck Observatory, which is operated as a scientific partnership among the California Institute of Technology, the University of California and the National Aeronautics and Space Administration. The Observatory was made possible by the generous financial support of the W. M. Keck Foundation.
    The authors wish to recognize and acknowledge the very significant cultural role and reverence that the summit of Maunakea has always had within the indigenous Hawaiian community. We are most fortunate to have the opportunity to conduct observations from this mountain. Finally, this research has made use of the Keck Observatory Archive (KOA), which is operated by the W.M. Keck Observatory and the NASA Exoplanet Science Institute (NExScI), under contract with the National Aeronautics and Space Administration.
    
    This work was performed using the Cambridge Service for Data Driven Discovery (CSD3), part of which is operated by the University of Cambridge Research Computing on behalf of the STFC DiRAC HPC Facility (\url{www.dirac.ac.uk}). The DiRAC component of CSD3 was funded by BEIS capital funding via STFC capital grants ST/P002307/1 and ST/R002452/1 and STFC operations grant ST/R00689X/1. This work further used the DiRAC@Durham facility managed by the Institute for Computational Cosmology on behalf of the STFC DiRAC HPC Facility. The equipment was funded by BEIS capital funding via STFC capital grants ST/P002293/1 and ST/R002371/1, Durham University and STFC operations grant ST/R000832/1. DiRAC is part of the National e-Infrastructure.

\facilities{Keck:II (ESI), VLT:Kueyen (X-Shooter)}

\software{
    {\tt Astrocook} \citep[][]{cupani_astrocook_2020}, 
    {\tt Astropy} \citep{astropy_collaboration_astropy_2013},
    {\tt Matplotlib} \citep{hunter_matplotlib_2007},
    {\tt NumPy} \citep{van_der_walt_numpy_2011},
    {\tt SpectRes} \citep{carnall_spectres_2017}
}
\end{acknowledgments}

\appendix
\section{Relationship Between \lyb\ Dark Gaps and \lya\ Dark Gaps} \label{app:Lyb_Lya}
\begin{figure*}[]
    \centering
    \includegraphics[width=6in]{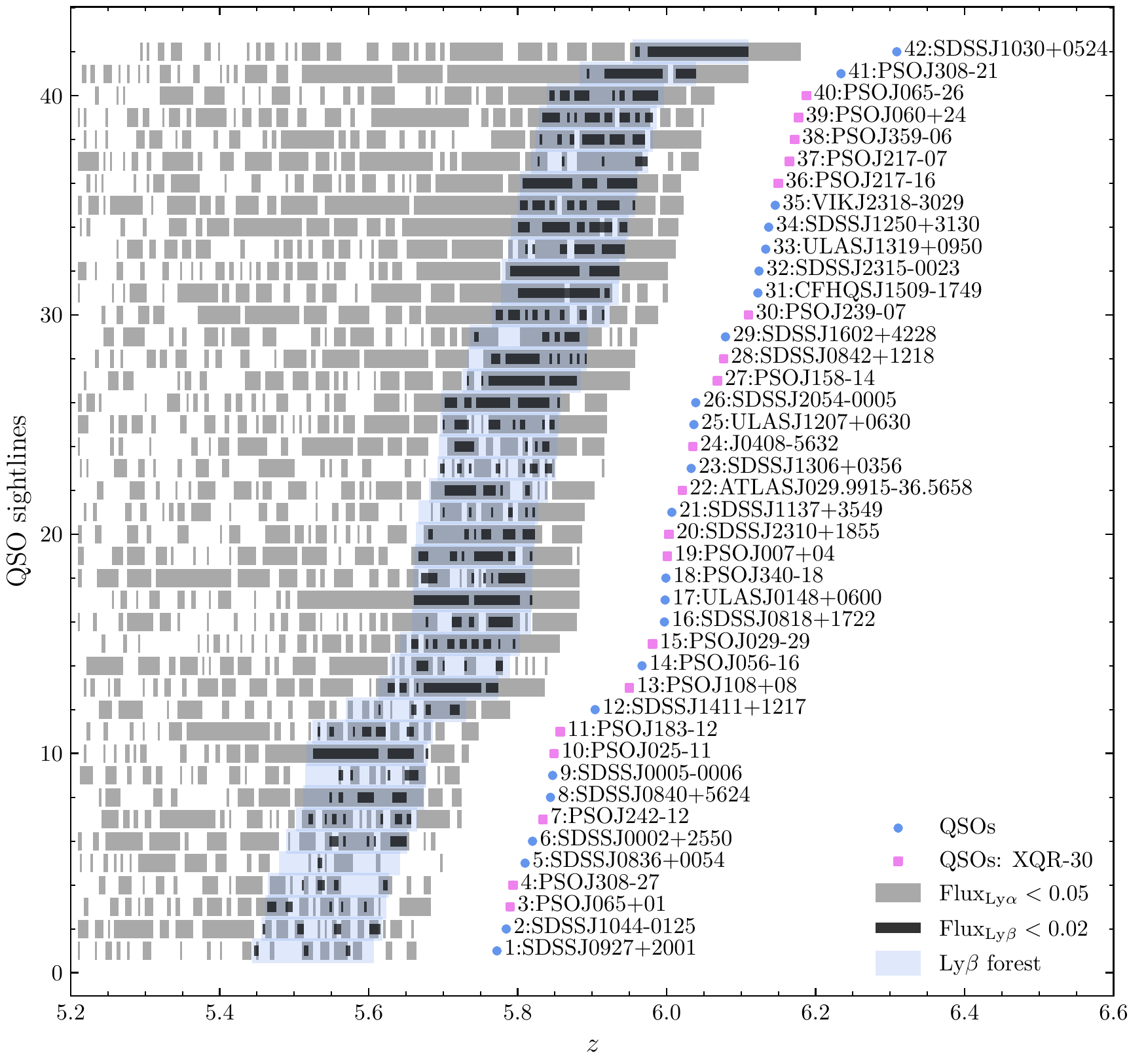}
    \caption{ Overview of \lyb-opaque regions and \lya\ dark gaps from our sample of 42 QSO lines of sight. Black bars show \lyb-opaque regions, where normalized flux in the \lyb\ forest $\rm Flux_{Ly\beta}<0.02$ per $1\cmpch$ bin. Gray bars show \lya\ dark gaps as defined in \citetalias{zhu_chasing_2021}. Light blue shades highlight the redshift ranges of the \lyb\ forest. The overlap between the gray bars and black bars yields \lyb\ dark gaps as defined in this work.  
    }
    \label{fig:all_los_lya}
\end{figure*}

To illustrate the effects of requiring \lyb\ gaps to also be dark in the \lya\ forest, here we explore the relationship between \lyb\ dark gaps and \lya\ dark gaps. In Figure \ref{fig:all_los_lya} we over-plot \lyb-opaque regions ($\rm Flux_{Ly\beta}<0.02$ per $1\cmpch$ bin) on \lya\ dark gaps as defined in \citetalias{zhu_chasing_2021}. Although \lyb-opaque regions overlap strongly with \lya\ dark gaps, there do exist regions that are dark only in the \lyb\ forest, e.g., the long \lyb-opaque region toward CFHQS J1509$-$1749 that bridges two \lya\ dark gaps, as shown in the figure. These cases are due to foregorund \lya\ absorption in the \lyb\ forest  Requiring \lyb\ gaps to also be dark in the \lya\ forest partially avoids this kind of foreground contamination.

\begin{figure*}[]
    \centering
    \includegraphics[width=3.5in]{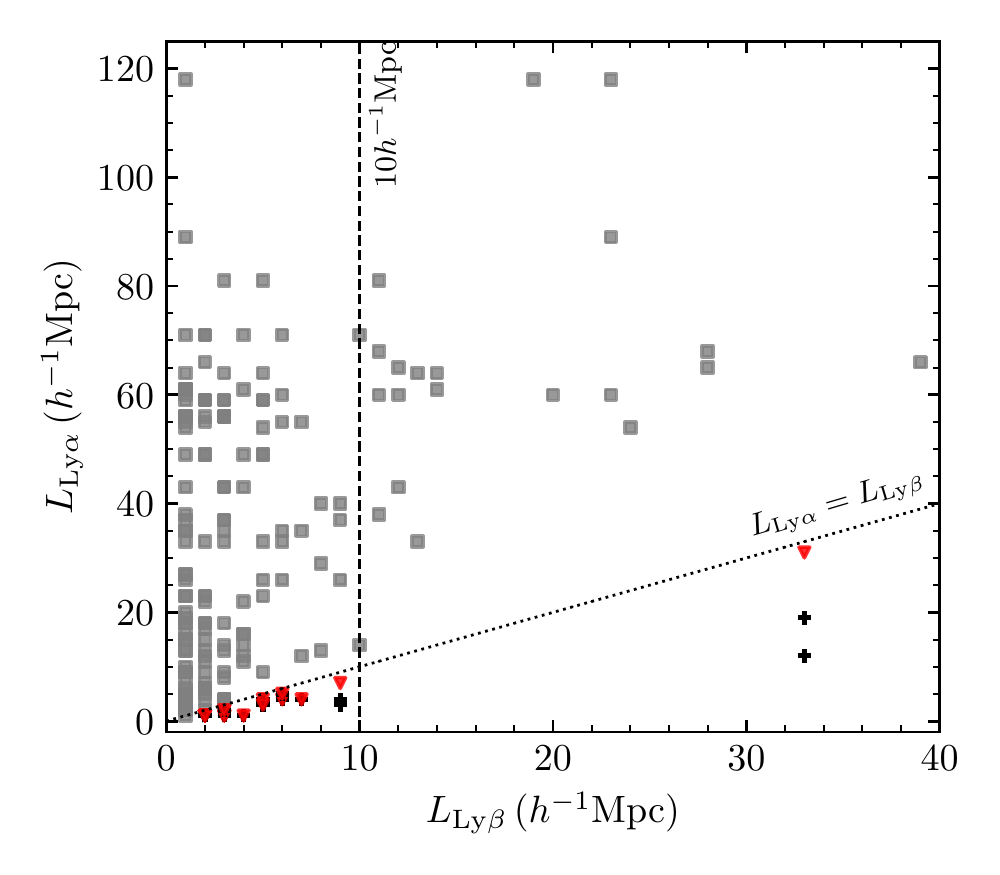}
    \caption{ Length of \lya\ dark gaps versus length of \lyb-opaque regions. For \lyb\ dark gaps that are entirely within the redshift range of the \lya\ dark gap, we plot the length of the \lya\ gap versus the length of the \lyb\ gap with a gray square. Red triangles denote situations where not all $1\cmpch$ pixels in a \lyb-opaque region have $\rm Flux_{Ly\alpha}<0.05$ (\lya\ dark gaps). The path length of \lya\ dark gaps inside these \lyb-opaque regions are marked with black crosses.
    }
    \label{fig:LLya_LLyb}
\end{figure*}

We further plot the length of \lya\ dark gaps versus the length of corresponding \lyb-opaque regions in Figure \ref{fig:LLya_LLyb}. Most of long ($\geq10\cmpch$) \lyb\ dark gaps appear in $L\geq 30 \cmpch$ \lya\ dark gaps. Only one out of 23 long \lyb-opaque regions contains transmission in \lya\ and is split into two \lyb\ dark gaps.

\section{Metal-Enriched Systems \label{app:moreMetal}}
\begin{figure*}[htb!]
    \centering
    \includegraphics[width=6in]{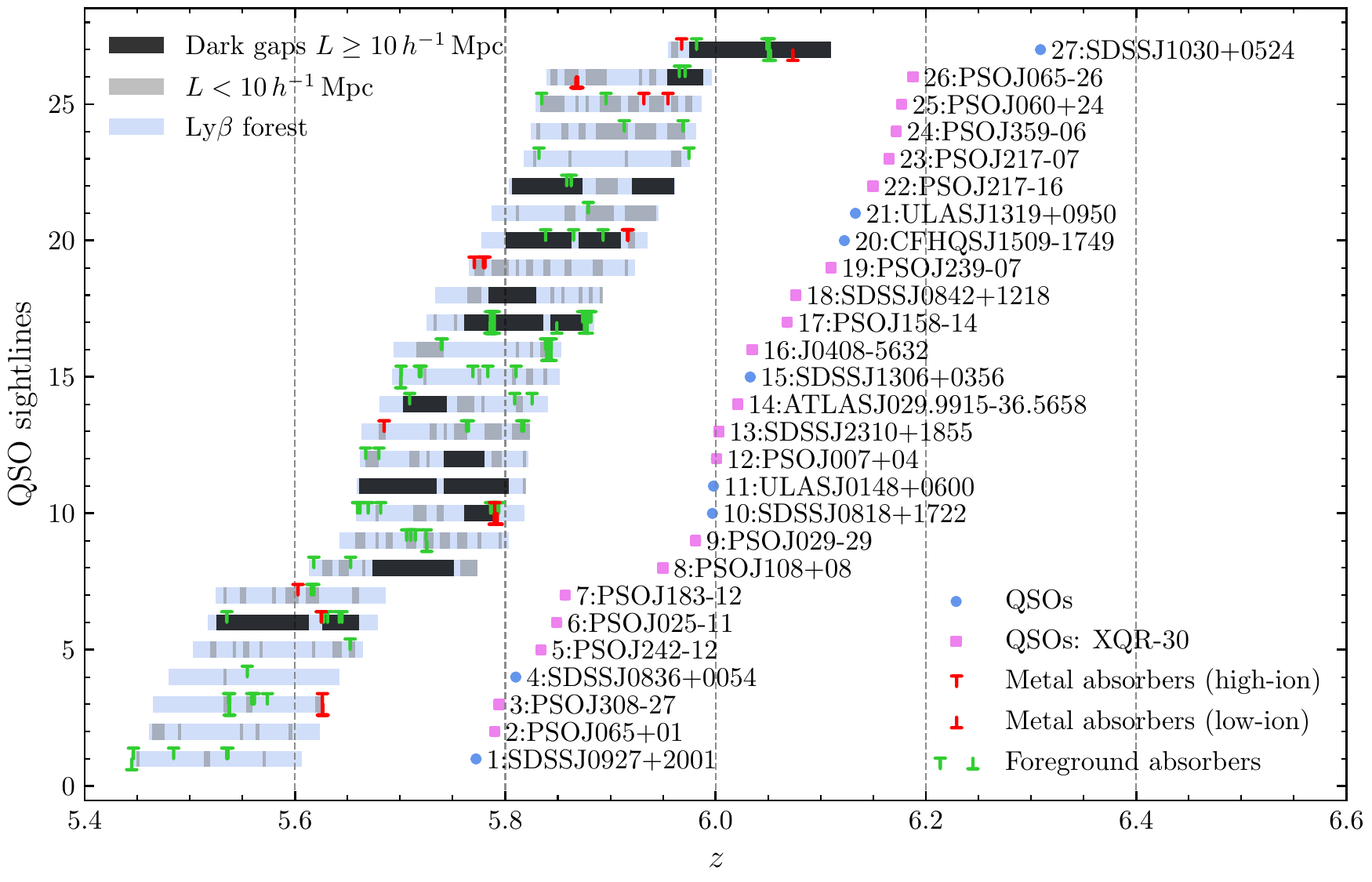}
    \caption{Similar to Figure~\ref{fig:all_los}, but with metal-enriched absorbers over-plotted for 27 sightlines that have a relatively complete and consistent identification of metal-enriched absorbers (Davies et al., in prep). We label high-ionization and low-ionization systems within the redshift of the \lyb\ forest with red ``$\top$'' and ``$\bot$'' symbols, respectively. 
    Foreground metal-enriched absorbers whose \lya\ absorption would fall within the \lyb\ forest are labeled with green symbols at the corresponding \lyb\ redshifts.}
    \label{fig:moreMetal}
\end{figure*}

In Figure \ref{fig:moreMetal} we display an overview of dark gaps with metal-enriched systems over-plotted for the 27 QSO sightlines in our sample where the identification of metals is relatively complete and consistent. We label metal systems with redshifts in the \lyb\ forest and in the foreground \lya\ forest separately. These systems are included in a metal absorber catalog that will be presented in Davies et al. in prep. Briefly, the {\tt Python} application {\tt Astrocook} was used to perform an automated search for \ion{Mg}{2}, \ion{Fe}{2}, \ion{C}{4}, \ion{Si}{4}, and \ion{N}{5} absorbers, and DLA-like systems probed by \ion{C}{2} and other low-ionization species. Candidate absorbers were identified using a cross-correlation algorithm within {\tt Astrocook} that searches for redshifts where significant absorption is present in all relevant transitions. Custom filtering algorithms and visual inspection were then used to remove false positives and produce the final absorber list.

We then investigate the correlation between long ($L\geq 10 \cmpch$) dark gaps and metal systems.  We find that the probability for a metal system in the \lyb\ forest to lie in a long dark gap is $15 \pm 9\%$, where the $68\%$ confidence limit comes from bootstraping these 27 sightlines 10,000 times. This probability is $31 \pm 9\%$ in the case of a system in the corresponding foreground \lya\ forest.  In these calculations we count clustered metal absorbers with a separation of $<1\cmpch$ as one system.  By comparison, the probability that a randomly chosen point lies in a long dark gap is $22 \pm 9\%$.  Our results suggest that the correlation between long dark gaps and (foreground) metal systems is not highly significant, at least for this sub-sample. The relatively lower probability of finding metal absorbers within the redshifts of long dark gaps nevertheless potentially favors the association between high IGM \lya\ opacities and galaxy underdensities \citep[see also][]{becker_evidence_2018,kashino_evidence_2020,christenson_constraints_2021}.

\section{Uncertainties in the fraction of QSO spectra showing dark gaps \label{app:F01}}
\begin{figure*}[htb!]
    \begin{center}
        \gridline{
        \fig{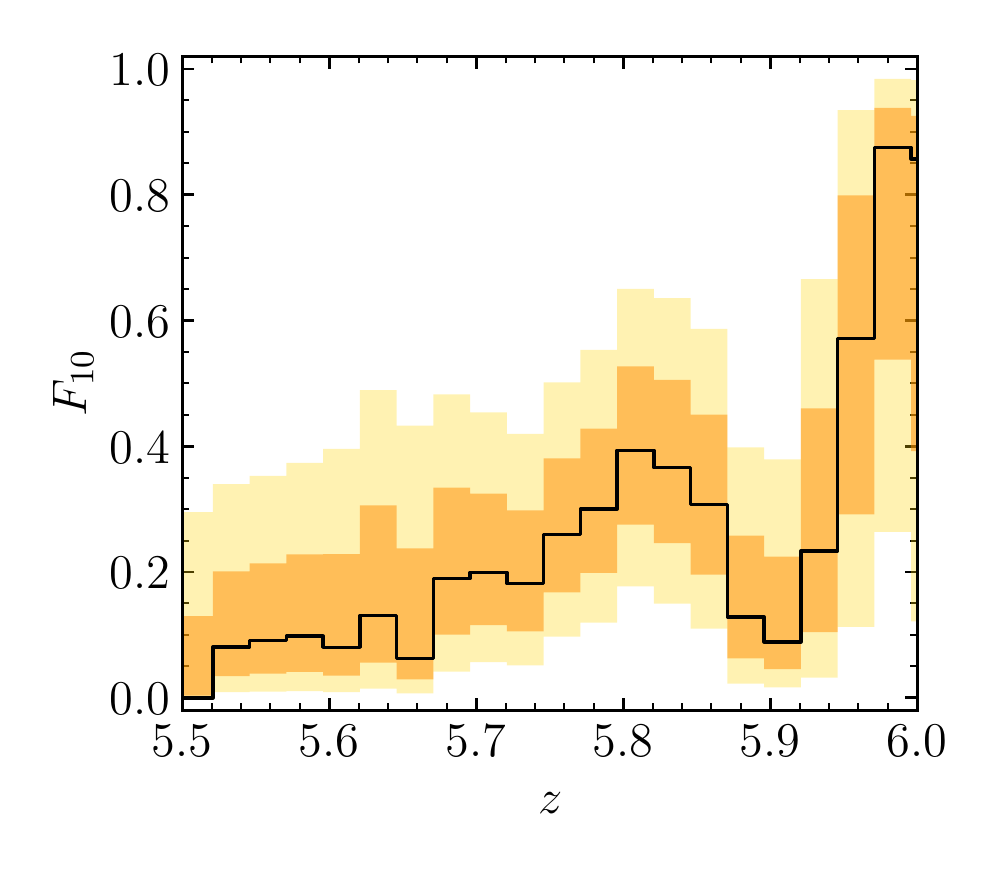}{2.9in}{(a)}
        \fig{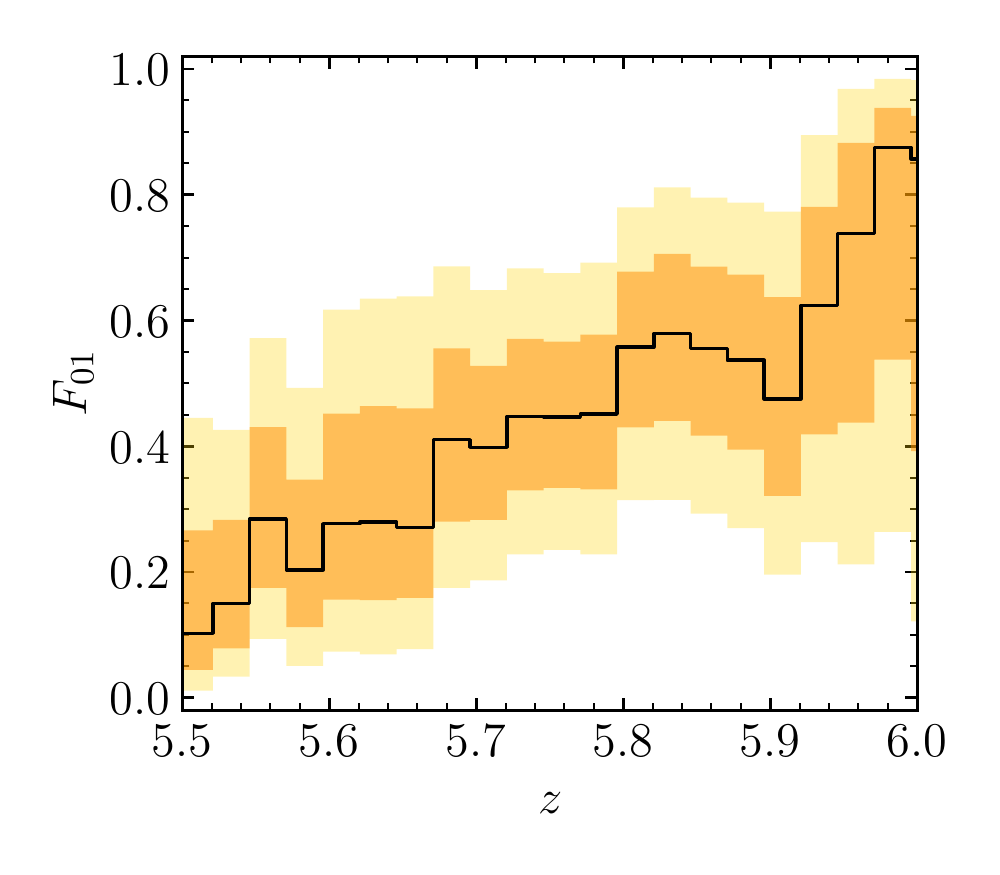}{2.9in}{(b)}
        }
        \caption{
        {\bf (a)} Statistical uncertainty estimation for $F_{10}$ shown in Figure \ref{fig:obs} (c). Dark and light shaded regions mark the 68\% and 95\% limits of $F_{10}$ based on Beta distribution. 
        {\bf (b)} Fraction of QSO spectra showing dark gaps with $L\geq 1\cmpch$.
        }
        \label{fig:F01}
    \end{center}
\end{figure*}

The evolution in $F_{10}$ shown in Figure~\ref{fig:F10} shows a large drop near $z = 5.9$. To estimate the statistical fluctuations in $F_{10}$, we treat the ``hit rate'' of long dark gaps at individual redshifts as a binomial experiment defined by the number of hits (number of long dark gaps, $n_{\rm dark}$) inside a different number of trials (number of QSO sightlines, $n_{\rm qso}$). At a certain redshift, the posterior probability distribution function for the true ``hit rate'', $x$, can be expressed as a Beta distribution, $f(x; \alpha, \beta)\propto x^{\alpha-1}(1-x)^{\beta-1} $, with $\alpha = n_{\rm dark}+0.5 $ and $\beta = n_{\rm qso}-n_{\rm dark}+0.5$, assuming a Jeffreys' prior. As shown in Figure \ref{fig:F01} (a), the evolution of $F_{10}$ is consistent with a monotonic increase with $z$ within the 95\% confidence intervals. We caution that the analysis here assumes that the ``hit rates'' at different redshifts are independent from each other. 

While the dip could be due to statistical fluctuations, we nevertheless wish to check whether it may relate to possible biases in the data related to \lyb\ absorption near that redshift. To check for possible systematic effects, we calculate the fraction of QSO spectra showing dark gaps of any length ($L\geq1\cmpch$) as a function of redshift, $F_{01}$. As shown in Figure~\ref{fig:F01}, the drop in $F_{10}$ at $z\sim5.9$ is not present in $F_{01}$. Instead, the evolution of \lyb-opaque regions with redshift appears relatively smooth. We thus find no evidence of systematic effects in the data that would suggest lower absorption overall near $z = 5.9$.

\section{Dark Gaps in a Lower-redshift Sample \label{app:lowz}}

Here we examine the extent to which strong, clustered absorbers associated with galaxies may be able to produce long dark gaps in the \lyb\ forest.  These (typically metal-enriched) absorbers may produce discrete absorption in either \lyb\ over the redshift over the trough or \lya\ at the corresponding foreground redshifts. They may also connect otherwise short dark gaps to form longer gaps. Of particular interest are very long gaps analogous to the $L=28\cmpch$ gap toward PSO J025$-$11.  To tests whether such gaps may be due to (circum-)galactic absorbers rather than the IGM, we search for dark gaps at $z \lesssim 5.5$ in a sample of QSO lines of sight that lie at somewhat lower redshifts than our main sample.  Because the IGM becomes increasingly transparent toward lower redshifts, any long dark gaps in this sample might signal a significant contribution from discrete systems associated with galaxies.

Our lower-redshift sample includes 27 ESI and X-Shooter spectra of QSOs over $5.0 < z_{\rm em} < 5.7$ from the Keck and VLT archives. 
The selection of targets is based on their redshift and is independent from foreknowledge of dark gaps. QSO spectra in this lower-redshift sample have S/N greater than 20 per pixel in the \lyb\ forest. 
In order to account for the increased mean transmission at low redshifts, we conservatively use a higher flux threshold of 0.08 when searching for dark gaps. The ratio of mean flux in the \lyb\ forest at $z=4.8$ and 5.6 is about 3.2  \citep[e.g.,][]{fan_constraining_2006,eilers_anomaly_2019,bosman_comparison_2021}, thus a flux threshold of 4 times the high-redshift value is used.

\begin{figure*}
    \centering
    \includegraphics[width=6in]{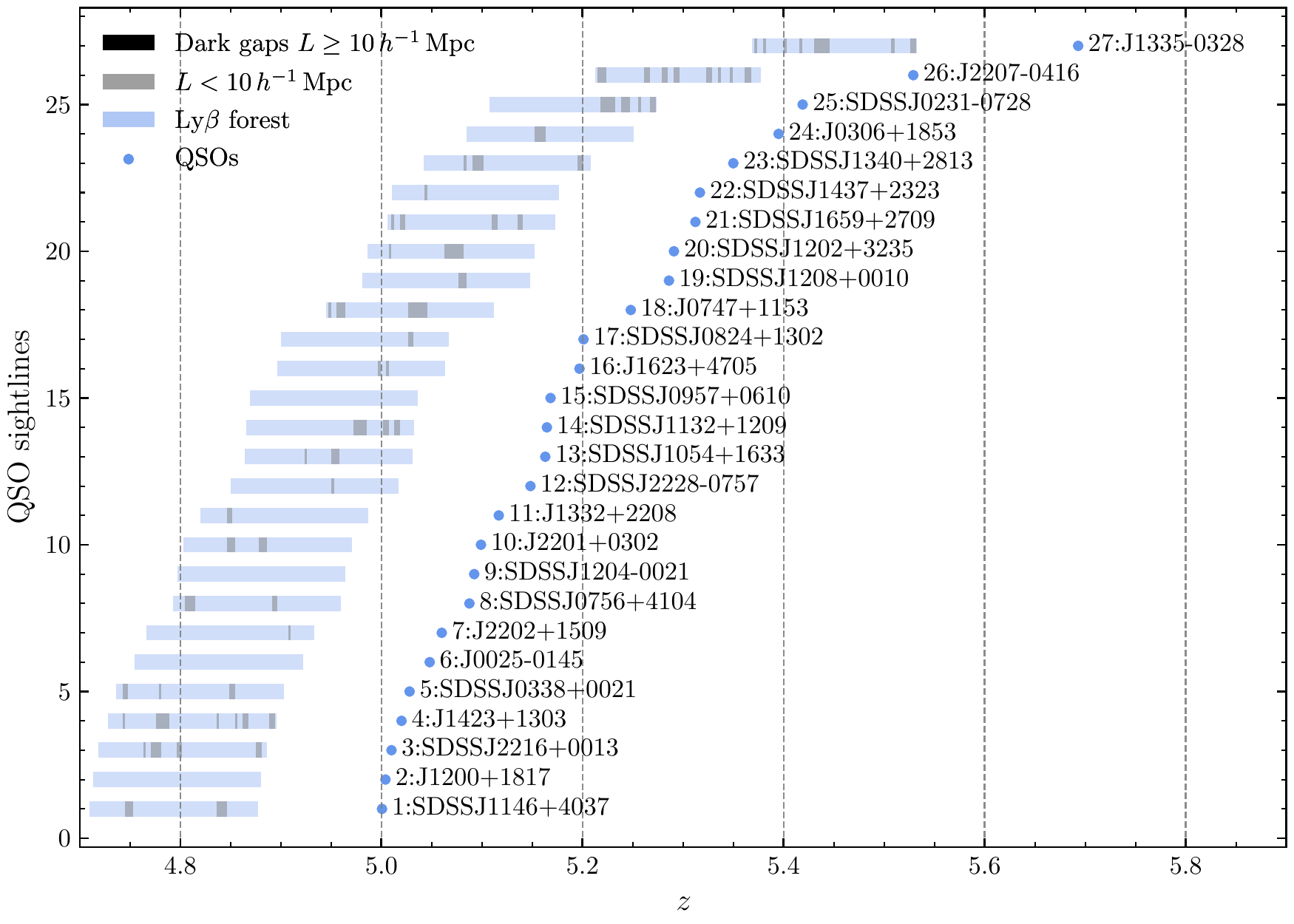}
    \caption{Similar to Figure~\ref{fig:all_los}, but showing dark gaps identified in the \lyb\ forest from a lower-redshift sample. 
    \label{fig:all_los_lowz}}
\end{figure*}

Figure~\ref{fig:all_los_lowz} shows dark gaps detected in this lower-redshift sample. No dark gaps longer than $10 \cmpch$ are detected.  
The lack of any long gaps in this sample suggests that extended gaps created largely by strong, discrete absorbers are rare, at least over $5 \lesssim z \lesssim 5.5$, 
which is reasonably close in redshift to our main sample.
This increases our confidence that the $L=28\cmpch$ dark gap toward PSO J025$-$11 is likely to mainly arise from IGM absorption. 

\pagebreak[2]

%
%
%
%
%

%

\end{document}